# MOCVD growth and band offsets of κ-phase $Ga_2O_3$ on c-plane sapphire, GaN- and AlN-on-sapphire and (100) YSZ substrates

A F M Anhar Uddin Bhuiyan[1,a)], Zixuan Feng[1], Hsien-Lien Huang[2], Lingyu Meng[1], Jinwoo Hwang[2], and Hongping Zhao[1,2,b)]

[1]*Department of Electrical and Computer Engineering, The Ohio State University, Columbus, OH 43210, USA*

[2]*Department of Materials Science and Engineering, The Ohio State University, Columbus, OH 43210, USA*

[a)] Email: bhuiyan.13@osu.edu [b)]Corresponding author Email: zhao.2592@osu.edu

## Abstract

Epitaxial growth of κ-phase $Ga_2O_3$ thin films are investigated on c-plane sapphire, GaN- and AlN-on-sapphire, and (100) oriented yttria stabilized zirconia (YSZ) substrates via metalorganic chemical vapor deposition (MOCVD). The structural and surface morphological properties are investigated by comprehensive material characterization. Phase pure κ-$Ga_2O_3$ films are successfully grown on GaN-, AlN-on sapphire, and YSZ substrates through a systematical tuning of the growth parameters including the precursor molar flow rates, chamber pressure and growth temperature, whereas the growth on c-sapphire substrates leads to a mixture of β- and κ-polymorphs of $Ga_2O_3$ under the investigated growth conditions. The influence of the crystalline structure, surface morphology and roughness of κ-$Ga_2O_3$ films grown on different substrates are investigated as a function of precursor flow rate. High resolution scanning transmission electron microscopy (HR-STEM) imaging of κ-$Ga_2O_3$ films reveals abrupt interfaces between the epitaxial film and the sapphire, GaN and YSZ substrates. The growth of single crystal orthorhombic κ-$Ga_2O_3$ films is confirmed by analyzing the STEM nano-diffraction pattern. The chemical composition, surface stoichiometry, and the bandgap energies of κ-$Ga_2O_3$ thin films grown on different substrates are studied by high resolution x-ray photoelectron spectroscopy (XPS) measurements. The type-II (staggered) band alignments at three interfaces between κ-$Ga_2O_3$ and c-sapphire, AlN, and YSZ substrates are determined by XPS, with the exception of κ-$Ga_2O_3$/GaN interface, which shows type I (straddling) band alignment.

## I. Introduction

Gallium oxide ($Ga_2O_3$) with its ultrawide bandgap of ~ 4.5-5.3 eV has recently emerged as a promising semiconductor material due its predicted high breakdown field strength (8 MV/cm), controllable n-type doping, availability of single crystal high quality native substrate and its capability for bandgap engineering by alloying with $Al_2O_3$ and $In_2O_3$ [1-33]. Among the five known polymorphs of $Ga_2O_3$ (α, β, γ, δ and κ-phases) [34], majority of the research have been focused on the epitaxial growth and development of high-performance electronic devices based on β-phase $Ga_2O_3$ due to its availability of high quality single crystal, large native substrates, excellent electrical properties, higher chemical and thermal stability. While heteroepitaxial growth of high-quality β-$Ga_2O_3$ films on symmetric hetero-substrates is challenging due to its monoclinic structure, κ-phase $Ga_2O_3$ (also refers as ε-phase) with a bandgap energy similar to β-$Ga_2O_3$ [35, 36] potentially allows for the growth of high-quality films on commonly used hexagonal hetero-substrates including GaN, AlN, SiC or sapphire. κ-$Ga_2O_3$ has an orthorhombic crystal structure with space group $Pna2_1$ [35, 36]. According to first-principles calculations, the κ-phase is the second most stable polymorph after β-phase $Ga_2O_3$ [36]. Furthermore, κ-$Ga_2O_3$ has strong spontaneous electrical polarization and ferroelectric characteristics [37, 38]. The spontaneous polarization of κ-$Ga_2O_3$ (0.31 $Cm^{-2}$ along its c-direction) is predicted to be much higher than that of group III-nitride semiconductors such as GaN (-0.034 $Cm^{-2}$) and AlN (-0.09 $Cm^{-2}$) [39], indicating the opportunity to generate a high-density of two-dimensional electron gases (2DEGs) without extrinsic doping. According to a recent theoretical study based on the density functional theory, the spontaneous polarization of κ-$Ga_2O_3$ can achieve a 2DEG density up to $10^{14}$ $cm^{-2}$ at κ-$Ga_2O_3$/m-AlN (m-GaN) heterointerfaces without additional doping, which is two orders of magnitude higher than the conventional AlGaN/GaN heterojunctions [39], potentially allowing the fabrication of κ-$Ga_2O_3$ based high-electron-mobility transistors (HEMTs) for high-power and high-frequency electronic applications.

Several investigations on the growth of κ-$Ga_2O_3$ have been performed. High quality heteroepitaxial growth of binary κ-$Ga_2O_3$ has been demonstrated by using different growth techniques including halide vapor phase epitaxy (HVPE) [40, 41], atomic layer deposition (ALD) [42], metal organic chemical vapor deposition (MOCVD) [35, 42-46], mist CVD [47-48], molecular beam epitaxy (MBE) [49, 50], and pulsed laser deposition (PLD) [51] on different foreign substrates, such as (0001) AlN [40], (0001) GaN [40], (111) YSZ [48], (111) MgO [48], (0001) $Al_2O_3$ [42, 44, 46, 49], (001) and (111) 3C-SiC [42] and (001) 6H-SiC [43]. In addition, ternary alloys with Al [52, 53] or In [54] were also investigated using PLD [52] and mist-CVD [53, 54] growth methods. Si doped electrically conductive κ-$Ga_2O_3$ thin films have been recently reported by MOCVD [55]. Tin (Sn) and indium (In) were found to facilitate the growth of κ-$Ga_2O_3$ films in PLD [51], MBE [49, 50] and mist-CVD [56] growth techniques. Coherent growth of orthorhombic $Ga_2O_3/(Al_xGa_{1-x})_2O_3$ superlattice structures were demonstrated by Sn-assisted PLD growth technique [57].

In 1952, the first to disclose this polymorph, Roy et al., were unable to determine the crystalline structure of ε/κ-$Ga_2O_3$ but inferred that its symmetry might be less than hexagonal or tetragonal [34]. Later in 2013, Playford et al. used neutron diffraction to investigate it further and discovered that it has a hexagonal structure identical to 2H-GaN with space group $P6_3mc$ [58].

According to recent experimental findings, κ-$Ga_2O_3$ possesses an orthorhombic crystal structure [49, 59, 60]. The hexagonal symmetry found in ε-phase has been explained by the twinning of three rotational domains of the orthorhombic κ-$Ga_2O_3$, revealing that the hexagonal $P6_3mc$ structure as proposed by Playford et al. [58] consists of orthorhombic $Ga_2O_3$ domains with $Pna2_1$ space group symmetry [49].

The crystal structure of selected substrate was found to be one of the important factors determining the polymorphic composition and structure of the epi-film. Epitaxial growth of κ/ε-$Ga_2O_3$ on hexagonal substrates such as AlN, GaN, and SiC, for example, resulted in the same hexagonal ε-$Ga_2O_3$ growth with $P6_3mc$ structure [40, 42, 43]. Aside from the crystal structure of the substrates, the growth conditions such as growth temperature, chamber pressure, precursor flow rates, and the VI/III ratio can also have a significant influence on the phase stabilization of $Ga_2O_3$. For example, a faster growth rate in HVPE growth method was found to facilitate the formation of phase pure κ-$Ga_2O_3$ [40]. Low growth temperature, on the other hand, was required for the deposition of κ-$Ga_2O_3$ but not a sufficient growth condition [40]. Although several studies on the epitaxial development of κ-$Ga_2O_3$ thin films have been conducted, the systematical growth mapping of κ-$Ga_2O_3$ films grown by MOCVD on various substrates are still limited. Given the promise of κ-$Ga_2O_3$ in the development of high-electron-mobility transistors due to its potential in forming 2DEGs, extraction of the band offsets is critical for device design and quantitative study of carrier confinement at the interface. The band discontinuities at the heterointerfaces of κ-$Ga_2O_3$ and foreign substrates are not reported yet.

In this work, we investigated the influence of different substrates on the epitaxial stabilization of κ-$Ga_2O_3$ thin films by using a number of foreign substrates including sapphire, AlN, GaN, and YSZ by MOCVD growth technique. The structural and morphological properties of κ-$Ga_2O_3$ are evaluated by comprehensive characterization via x-ray diffraction (XRD), high resolution x-ray photoelectron spectroscopy (XPS), atomic force microscopy (AFM), field emission scanning electron microscopy (FESEM) and high-resolution STEM and nanodiffraction. While previously the MOCVD growth of κ-$Ga_2O_3$ have been demonstrated on sapphire, SiC and GaN substrates [35, 42-46], the systematic study on MOCVD growth mapping by varying growth parameters is limited. MOCVD growth of κ-$Ga_2O_3$ thin films on c-sapphire substrates have been demonstrated with transition layers containing different phases of $Ga_2O_3$ (α, β and γ-phases) [46], whereas our STEM imaging on c-sapphire substrates showed high quality epitaxial growth of κ-$Ga_2O_3$ films without any transition layer of different phases. Although PLD and mist-CVD grown κ-$Ga_2O_3$ thin films on (111) oriented YSZ substrates are demonstrated [48, 51], we investigated a new orientation of YSZ substrate for the epitaxial growth of κ-$Ga_2O_3$ film by MOCVD, which reveals phase pure κ-$Ga_2O_3$ thin film growths. The structural and surface morphological properties and phase stabilization of κ-$Ga_2O_3$ films grown on all four substrates (c-plane sapphire, AlN, GaN, and YSZ) is studied by systematically varying growth conditions, such as the chamber pressure, temperature, VI/III ratio, growth rates and film thickness. The investigation of interfacial quality between κ-$Ga_2O_3$ and different substrates by atomic resolution STEM images reveals strong influence of underlying substrates. The chemical composition, surface stoichiometry and the band offsets are also determined at κ-$Ga_2O_3$/GaN (AlN, YSZ and $Al_2O_3$) heterointerfaces by utilizing XPS.

## II. Experimental Section

κ-$Ga_2O_3$ thin films were grown in Agnitron Agilis MOCVD reactor on c-plane (0001) GaN-on-sapphire, c-plane (0001) AlN-on-sapphire, (100) YSZ and c-plane sapphire [(0001) $Al_2O_3$] substrates. Triethylgallium (TEGa) and pure $O_2$ gas were used as Ga and O precursors, respectively. Argon (Ar) was used as carrier gas. The growth temperature and reactor pressure were varied between 510 - 650 °C and 4 - 60 torr, respectively. TEGa molar flow rates were varied from 19.12 to 52.57 μmol/min. $O_2$ flow rate was kept at 500 sccm. The growth rate was in the range of 1.2-4.85 nm/min. The substrates were cleaned ex-situ with solvents before loaded into the growth chamber.

XRD measurements by using a Bruker D8 Discover (Cu Kα radiation x-ray source, λ =1.5418 Å) were performed to evaluate the crystalline structure and quality of the films. XRD asymmetrical reciprocal space mapping was used to analyze the strain of the films grown on different substrates. Surface morphology and roughness were evaluated by using FESEM (FEI Helios 600) and AFM (Bruker AXS Dimension Icon), respectively. The film thicknesses were determined using cross-sectional FESEM and STEM High angle annular dark field (HAADF) images. The crystalline purity and the bandgap of κ-$Ga_2O_3$ thin films grown on different substrates were determined by performing XPS measurements (Kratos Axis Ultra X-ray photoelectron spectrometer with a monochromatized Al Kα x-ray source, $E_{photon}$ = 1486.6 eV). The band offsets at the heterointerfaces of κ-$Ga_2O_3$ and substrate were measured by utilizing XPS with an energy resolution of 0.1 eV. For high resolution scans, the electron pass energy was set to 20 eV, while for survey scans, it was set to 80 eV. The binding energy was calibrated using the C 1s core level at 284.8 eV. HAADF STEM images were obtained using a Thermo Fisher Scientific Titan scanning transmission electron microscope, operating at 300 kV.

**III. Results and Discussions**

The crystalline structure, quality, and orientation of the κ-$Ga_2O_3$ thin films grown on different substrates are analyzed by high resolution XRD. The films were grown with an optimized growth temperature at 620 °C and a chamber pressure of 5 torr. Figures 1 (a)-(d) show the XRD

ω-2θ scans of the κ-$Ga_2O_3$ films grown on (100) YSZ, (0001) AlN- and GaN-on-sapphire templates, and c-sapphire substrates, respectively. The (0006) α-$Al_2O_3$, (0002) GaN and AlN and (200) YSZ diffraction peaks are originated from the substrates. High intensity distinguishable diffraction peaks from (004) reflections of κ-$Ga_2O_3$ are observed for all the films grown on various substrates. Growth of phase pure κ-$Ga_2O_3$ films are obtained on GaN, AlN and YSZ substrates by optimizing the growth conditions (discussed in later sections), as indicated by the single and high intensity (004) κ-$Ga_2O_3$ diffraction peaks located at 2θ ≈ 38.8° (Figs. 1(a)-(c)). No additional peaks corresponding to other polymorphs of $Ga_2O_3$ are observed. However, on c-sapphire substrate, as shown in Fig. 1(d), ($\bar{4}02$) β-phase $Ga_2O_3$ peak is observed along with (004) κ-$Ga_2O_3$ peak, revealing that the crystalline phase of $Ga_2O_3$ is strongly influenced by the underlying substrates. XRD rocking curve full width at half maximum (FWHMs) of 0.64°, 2.19°, 0.81° and 1.09° are obtained from (004) reflection of κ-$Ga_2O_3$ films grown c-sapphire, AlN- and GaN-on-sapphire and YSZ substrates, respectively, indicating a strong influence of substrates on the structural quality of epitaxial κ-$Ga_2O_3$ thin films.

Asymmetric reciprocal space mapping (RSM) was performed to investigate the strain in ~200 nm thick κ-$Ga_2O_3$ films grown on c-sapphire substrate and GaN-on-sapphire template. Figures 2(a) and (b) show the asymmetric RSMs of these two samples in the vicinity of (139) κ-$Ga_2O_3$, (11$\bar{2}$.12) sapphire and (10$\bar{1}$5) GaN reflections. The (11$\bar{2}$.12) and (10$\bar{1}$5) reflections are originated from the sapphire and GaN substrates, respectively. While the vertical alignment of the substrate and epilayer peak positions at same $Q_x$ value indicates the epitaxial growth of pseudomorphically strained film, the maximum reflection intensity of (139) κ-$Ga_2O_3$ for both samples grown on c-sapphire substrate (Fig. 2(a)) and GaN-on-sapphire template (Fig. 2(b)) are found to be shifted from the fully strained line located at the substrate peak positions (parallel to out-of-plane reciprocal space lattice constant, $Q_z$). This characteristic corresponds to the fully relaxed epitaxial growth of ~200 nm thick κ-$Ga_2O_3$ films on both sapphire and GaN substrates owing to their large lattice mismatches with the substrates.

The epitaxial relationship between the κ-$Ga_2O_3$ thin films and different substrates was further investigated by XRD φ-scanning. Figures 3(a)-(d) show the XRD φ-scan profiles of the asymmetric {139} plane of κ-$Ga_2O_3$ thin films and $\{10\bar{1}2\}$ $Al_2O_3$, $\{10\bar{1}5\}$ GaN, $\{10\bar{1}5\}$ AlN and {311} YSZ planes of the substrates, respectively. For the $\{10\bar{1}2\}$ reflex from sapphire substrate ($2\theta \approx 25.6°$ and $\chi \approx 57.61°$), threefold symmetry is observed as evidenced by three diffraction peaks appearing every 120° (Fig. 3(a)). On the other hand, six-fold symmetry with a separation of 60° is observed for both $\{10\bar{1}5\}$ reflections of hexagonal GaN ($2\theta \approx 105.11°$ and $\chi \approx 20.58°$) and AlN ($2\theta \approx 111.22°$ and $\chi \approx 20.29°$) substrates. Four peaks appearing every 90°, in case of {311} YSZ ($2\theta \approx 59.63°$ and $\chi \approx 25.23°$) reveal four-fold symmetry of YSZ substrates owing to its cubic crystal structure. The φ-scan of {139} planes ($2\theta \approx 107.76°$ and $\chi \approx 22.30°$) of the orthorhombic κ-$Ga_2O_3$ thin films in Figs 3(a-c) show six asymmetric distinguishable diffraction peaks separated by 60°, indicating a six-fold symmetry of {139} κ-$Ga_2O_3$ with three in-plane rotational domains rotated by an angle of 120°. Owing to the orthorhombic structure of κ-$Ga_2O_3$, these three rotational domains have also been observed in previous studies [40, 43, 47, 54]. Due to the difference between the crystal structure of the substrates and epi-films and large lattice mismatches, the orthorhombic arrangement of κ-$Ga_2O_3$ leads to the rotation of in-plane domains. The in-plane and out-of-plane epitaxial relationships, based on the position of the substrate peaks with respect to those of the thin films, could be determined as (001) κ-$Ga_2O_3$ [130] ∥ (0001) GaN $[10\bar{1}0]$ and (001) κ-$Ga_2O_3$ [130] ∥ (0001) AlN $[10\bar{1}0]$ for the films grown on GaN- and AlN-on-sapphire templates, respectively, which agree well with previous studies [53, 54]. By comparing the interplanar spacings of {130} κ-$Ga_2O_3$ with $\{10\bar{1}0\}$ AlN (GaN) planes, the in-plane lattice mismatches of -6.63% (-8.94%) are determined between κ-$Ga_2O_3$ and AlN. (GaN). Additionally, the epitaxial relationship between κ-$Ga_2O_3$ and c-plane sapphire could be determined as (001) κ-$Ga_2O_3$ [130] ∥ (0001) $Al_2O_3$ $[11\bar{2}0]$, revealing relatively smaller in-plane lattice mismatch of +5.11% between κ-$Ga_2O_3$ and c-sapphire as compared to GaN and AlN [54]. While three in-plane rotational domains are observed in κ-$Ga_2O_3$ films grown on c-sapphire, GaN and AlN substrates, the twelve asymmetric diffraction peaks corresponding to the {139} plane of κ-$Ga_2O_3$ grown on cubic (100) YSZ substrates reveal the twelve-fold symmetry with higher rotational domains as compared to other substrates (Fig. 3(d)). Previous studies on growth of κ-$Ga_2O_3$ thin films were demonstrated on (111) oriented YSZ substrates by mist-CVD [48] and PLD [51]. Only six-fold symmetric spots corresponding to the $\{10\bar{1}4\}$ planes of κ-$Ga_2O_3$ and three-fold symmetric spots corresponding to the {200} planes of the YSZ substrate were observed in XRD pole figure in mist-CVD grown κ-$Ga_2O_3$ films [48]. Similar to mist-CVD, the PLD grown κ-$Ga_2O_3$ films on (111) YSZ substrates showed three-fold symmetry for {200} planes of YSZ and six fold symmetry for {131} and {206} planes of κ-$Ga_2O_3$ from XRD φ-scan, whereas 12 peaks were observed for {122} and {212} lattice planes [51], confirming the orthorhombic structure of the crystal lattice. While previous reports demonstrated the growth of κ-$Ga_2O_3$ films on (111) oriented YSZ substrates, the investigation of the growth of κ-$Ga_2O_3$ film on (100) oriented YSZ substrates revealed 12 peaks corresponding to {139} planes of κ-$Ga_2O_3$, indicating 6 in-plane rotational domains due to the mismatch of rotational symmetry between the cubic YSZ substrate and orthorhombic κ-$Ga_2O_3$ epilayer. Typically, the rotational domains are originated from the epitaxial growth of a crystalline material with a low degree of symmetry on a substrate with a high degree of symmetry, implying that the substrates with the same orthorhombic crystal structure and smaller lattice mismatch can potentially prevent the formation of in-plane rotation domains of κ-$Ga_2O_3$ films [61].

The surface morphology and roughness of κ-$Ga_2O_3$ films are evaluated by FESEM and AFM imaging. Figures 4 (a)-(d) show the surface FESEM images of ~ 200 nm thick κ-$Ga_2O_3$ films grown on c-sapphire, GaN- and AlN-on-sapphire templates and YSZ substrates, respectively. The films were grown with TEGa flow rate of 220 sccm at 620 °C and 5 torr chamber pressure. Smooth and uniform surface morphologies are observed for the films grown on sapphire, GaN and YSZ substrates. Despite of possessing a relatively smaller lattice mismatch as compared to GaN [39], κ-$Ga_2O_3$ films grown on AlN-on-sapphire template show rougher surface morphologies as shown in Fig. 4(c), which might be due to higher RMS surface roughness of AlN-on-sapphire template as discussed in next paragraph and surface reconstruction by the strong oxidation of Al adatoms on growth surface or decomposition of AlN under oxygen atmosphere. Previously, it was found that AlN bulk substrates with smooth surface morphologies start to oxidize between 800-900 °C [62], whereas the oxidation temperature can vary between 550 to 1100 °C for AlN powder [63]. However, in addition to the temperature, the oxidation of AlN might also depend on other factors, such as the duration of oxidation, $O_2$ precursor flow rates, chamber pressure etc. In previous MBE growth study, the decomposition of AlN was also strongly dependent on growth conditions (vacuum, nitrogen gas and plasmas) [64]. While significant levels of III-N decomposition occur even under optimized MBE growth conditions [64], the AlN-on-sapphire template with rougher surface morphology as shown in AFM image of Figure 5(c) is found to oxidize at our investigated growth conditions such as growth temperature of 550-650 °C, chamber pressure of 4-15 torr and $O_2$ flow of 500 sccm during the MOCVD growth of κ-$Ga_2O_3$ films under $O_2$ rich growth environment. The surface reconstruction by such oxidation of AlN may contribute to the roughening of interface quality as shown in HR-STEM imaging in later paragraphs.

To further probe the surface roughness of the films grown on different substrates, AFM imaging with a scan area of 5x5 $\mu m^2$ was performed on the κ-$Ga_2O_3$ films grown with TEGa flow rate of 80, 130 and 220 sccm as shown in Figures 5(a)-(p). All the films were grown for 40 mins of growth duration. With the increase of TEGa flow rates from 80 to 220 sccm (2.75x), the film thicknesses are found to increase from 48 to 194 nm (~ 4x). The nonlinear increase in the film thickness can be attributed to the pre-reaction of precursor in gas phase. The surface RMS roughness of κ-$Ga_2O_3$ films grown on different substrates are also compared with bare substrates as shown in Figures 5 (a)-(d). While c-sapphire, GaN-on-sapphire and YSZ bare substrates show lower RMS roughness, AlN-on-sapphire template shows rougher surface morphology. Among various substrates, κ-$Ga_2O_3$ grown on c-sapphire shows the lowest RMS roughness, ranging between 1.62 and 2.95 nm. While higher RMS roughness are observed for κ-$Ga_2O_3$ grown on AlN-on-sapphire templates for all TEGa flow rates, the RMS values, in case of c-sapphire, GaN-on-sapphire and YSZ substrates are found to increase with increasing TEGa flow rates, implying that the film growth rate and thickness strongly affect the surface morphology and roughness of κ-$Ga_2O_3$ films. The formation of aggregates due to the gas phase reaction of the precursor, as they reached to the growth surface, may also contributes to the roughening of the surface morphologies. The influence of film thickness on the surface morphologies of κ-$Ga_2O_3$ films grown with the same growth rate of 4.85 nm/min (TEGa flow rate: 220 sccm) is further investigated by AFM imaging, as shown in Figures 6(a)-(l). Using the same TEGa flow rate, three different κ-$Ga_2O_3$ films were grown with 20, 50 and 194 nm thickness on sapphire, GaN, AlN and YSZ substrates. With the increase of film thicknesses from 20 nm to 194 nm, the surface RMS roughness of κ-$Ga_2O_3$ films increases, indicating a strong influence of film thickness on the surface roughness of the epi-layer grown with similar growth rate.

The influence of different growth conditions such as the growth temperature and chamber pressure on the crystalline structure and quality of κ-$Ga_2O_3$ films grown on various substrates were further investigated. Figures 7(a)-(b) show the XRD ω-2θ scans of κ-$Ga_2O_3$ films grown on c-sapphire substrates at the growth temperature of 650 °C with different growth chamber pressures of 20 and 7 torr. The film grown at 20 torr chamber pressure leads to the growth of phase pure β-$Ga_2O_3$ thin films on c-sapphire (Fig. 7(a)), while reducing the pressure from 20 to 7 torr promotes the growth of both β- and κ- polymorphs of $Ga_2O_3$ as evidenced by the co-existence of ($\bar{2}01$) β-$Ga_2O_3$ and (001) κ-$Ga_2O_3$ as shown in Fig. 7(b). In addition, the chamber pressure was also varied from 20 to 60 torr to understand the influence of chamber pressure on the phase stability of $Ga_2O_3$ films grown on different foreign substrates including c-plane sapphire substrates, GaN- and AlN-on-sapphire and YSZ, as shown in Figure S1 of the supplementary material [78]. The films were grown at the growth temperature of 635 °C using TEGa flow of 220 sccm. While lower pressure (20 torr) promotes the mixture of κ and β phases of $Ga_2O_3$ films, increasing the pressure from 20 to 60 torr leads to the phase pure β-$Ga_2O_3$ films growths on all foreign substrates as shown in XRD ω-2θ scan spectra in Figure S1. Recent growth study on κ-$Ga_2O_3$ films on c-sapphire substrates have discussed the influence of growth rates on phase stabilization of κ-$Ga_2O_3$ [46], which revealed that the higher growth rates lead to phase pure κ-$Ga_2O_3$ film growths on c-sapphire substrates. Similarly, the nucleation and growth of κ- and β-polymorphs of $Ga_2O_3$ grown via MOCVD are also investigated on c-plane sapphire at temperatures ranging between 610 and 650 °C for different TMG and $H_2O$ flows using a horizontal reactor [65], resulting in a thickness (growth rate) variation due to the concentration gradient in the gas phase from the inlet toward the outlet. This study also indicated that the lower growth rates (i.e., lowering the supersaturation) facilitates the formation of β-$Ga_2O_3$ for both growth temperatures. The transformation from κ- to β-$Ga_2O_3$ occurred through a mixed-phase material for 650 °C, while an abrupt phase transition was observed at relatively lower temperature of 610 °C. For a given substrate temperature, the supersaturation creates severely non-equilibrium conditions at the inlet, whereas on the other side, the thermodynamic conditions are more favorable for nucleating the stable β-$Ga_2O_3$ as opposed to the metastable one. As long as the growth temperature is low, the lower activation barrier favors the nucleation of the metastable κ-$Ga_2O_3$. Similarly, in our study using far injection showerhead vertical reactor, we observed that lower pressure helps phase stabilization of κ-$Ga_2O_3$ thin films, which can be attributed to the higher growth rates of $Ga_2O_3$ films grown at lower chamber pressure condition. The growth rates increase as the gas phase reaction of precursors is suppressed at lower chamber pressure [3, 20], which promotes the phase stabilization of κ-$Ga_2O_3$. Although a lower growth pressure tends to facilitate the epitaxial growth of κ-$Ga_2O_3$, phase pure κ-$Ga_2O_3$ could not be achieved on c-sapphire substrates even with further tuning of the growth pressure from 3-20 torr (not shown) and growth temperature from 510-650 °C, indicating a narrow MOCVD growth window for κ-$Ga_2O_3$ on c-sapphire. The increase of the growth rates by further decreasing of the chamber pressure and/or increasing the Ga precursor flow rates might possibly stabilize κ-$Ga_2O_3$ on c-plane sapphire substrates. However, several other factors, such as the growth methods and conditions, selection of precursors, gas phase compositions and chemistry might also contribute to the phase stabilization of κ-$Ga_2O_3$ on c-sapphire substrates [66].

While the growth chamber pressure is found to be a key parameter influencing the crystalline structure of $Ga_2O_3$, the effect of growth temperature on the growth of κ-$Ga_2O_3$ films are also investigated for different substrates as shown in the XRD ω-2θ scans in Figures 8 (a)-(d). All the films were grown at a fixed reactor pressure of 5 torr, while the growth temperature was varied from 510 to 620 °C. For the $Ga_2O_3$ films grown on c-sapphire (Fig. 8(a)), the peak corresponding to (004) κ-$Ga_2O_3$ tends to dominate over ($\bar{4}$02) β-$Ga_2O_3$ peaks as the temperature increases from 510 to 620 °C, indicating that the optimal growth temperature for the MOCVD growth of κ-$Ga_2O_3$ is very narrow as further increasing the temperature induces β-phase $Ga_2O_3$ growth on c-sapphire. At relatively low temperature of 510 °C, no diffraction peaks corresponding to κ-$Ga_2O_3$ is found on AlN-on-sapphire and YSZ substrates and very low intensity peak of (004) κ-$Ga_2O_3$ along with a ($\bar{4}$02) β-$Ga_2O_3$ peak is observed for the growth on GaN-on-sapphire substrate (Fig. 8(b)-(d)). Nevertheless, phase pure κ-$Ga_2O_3$ thin films could be grown on (0001) GaN- and AlN-on-sapphire and (100) YSZ substrates at relatively higher growth temperatures of 550 and 620 °C, respectively, implying a relatively wider growth window in terms of growth temperature for κ-$Ga_2O_3$ on top of these substrates.

To understand the structural characteristics and growth mechanism of κ-$Ga_2O_3$ thin films, STEM and nanobeam electron diffraction analysis was performed, as shown in Fig. 9 (κ-$Ga_2O_3$ on sapphire), Fig. 10 (κ-$Ga_2O_3$ on GaN), Fig.11 (κ-$Ga_2O_3$ on AlN), and Fig.12 (κ-$Ga_2O_3$ on YSZ). As shown in Fig. 9a and Fig. 9b, the cross-sectional STEM HAADF and low-angle ADF (LAADF) images reveal columnar contrast, indicating a columnar growth of κ-$Ga_2O_3$ thin films grown on c-plane sapphire substrate. By comparing the experimental diffraction patterns to the simulated diffraction patterns (shown in inset), we identify the orthorhombic phase of κ-$Ga_2O_3$ with zone orientations of [100] and [110] for regions (i) and (ii) in Fig. 9a, respectively, and [010] α-$Al_2O_3$ (Fig. 9b) for the substrate (region (iii). Fig. 9c shows a magnified view of interface between κ-$Ga_2O_3$ layer and sapphire substrate. The contrast of the atomic columns in the κ-$Ga_2O_3$ film demonstrates the crystallographic relationship [49] where the 3 orthorhombic in-plane rotational domains (100), ($\bar{1}10$) and ($1\bar{1}0$) are superimposed along the *a*-axis projection. Furthermore, a resolved atomic column near the surface reveals a phase pure [100] κ-$Ga_2O_3$ (Figure 9d).

The structural quality of κ-$Ga_2O_3$ layers on (0001) GaN (Fig. 10), (0001) AlN (Fig. 11), and (100) YSZ (Fig. 12) substrates was examined using cross-sectional STEM HAADF and LAADF imaging. The electron diffraction results reveal the localized region with the zone orientation of [100] κ-$Ga_2O_3$ || [210] GaN (Fig. 10b), [310] κ-$Ga_2O_3$ || [100] AlN (Fig. 11b), and [010] κ-$Ga_2O_3$ || [$0\bar{1}2$] YSZ (Fig. 12b), respectively. In addition, the κ-$Ga_2O_3$ layer with rotational crystal domains are identified at the interface region (Fig. 10c, Fig. 11c, and Fig.12c). A high-resolution STEM HAADF image in a selected area also indicates a pure phase of [010] orthorhombic structure, confirming the epitaxial growth of (0001) κ-$Ga_2O_3$ on (100) YSZ substrate (Fig 12d). Despite of having rotational domains in κ-$Ga_2O_3$ layer, sharp and abrupt interfaces are observed between κ-$Ga_2O_3$ films and sapphire, GaN and YSZ substrates (Fig. 9c, Fig. 10c, and Fig. 12c). However, as a result of strong oxidation of Al adatoms on growth surface and decomposition of AlN under oxygen atmosphere, the interfacial sharpness between κ-$Ga_2O_3$ and AlN degrades (Figs. 11 (a, b)), resulting in a rougher surface morphology with higher RMS roughness values as compared to other substrates as shown in Fig. 4(c) and Figs. 5 (g, k, o).

Using high resolution XPS, the chemical compositions, purity, surface stoichiometry, band gaps, as well as the band offsets at the interface of κ-$Ga_2O_3$ and various substrates are investigated. Figure 13(a) shows the XPS survey spectra of κ-$Ga_2O_3$ films grown on c-sapphire, GaN- and AlN-on-sapphire and YSZ substrates, which indicates the growth of highly pure κ-$Ga_2O_3$ films as evidenced by the absence of any metallic contaminants in the spectra. Furthermore, the surface stoichiometry of κ-$Ga_2O_3$ films was also evaluated by determining the O/Ga ratio using O 1s and Ga 3s core level data. The Ga 3s and O 1s peaks are fitted with Gaussian-Lorentzian function after applying Shirley type background subtraction. The Ga 3s core level spectra is fitted with a single peak as shown exemplarily in Figure 13(b) for a representative κ-$Ga_2O_3$ thin films grown on YSZ substrate. The O 1s peak of the film is fitted with two components, O-Ga bond at 530.7 eV and O-H bond at 531.9 eV, as shown in Figure 13(c). The O/Ga ratio, determined by comparing the areas of Ga 3s and O 1s core levels after applying their corresponding sensitivity factors ($S_{Ga\ 3s}$ = 1.13, $S_{O\ 1s}$ = 2.93), are found to be 1.51, 1.51, 1.47 and 1.49 for κ-$Ga_2O_3$ films grown on c-sapphire, GaN- and AlN-on-sapphire and YSZ substrates respectively, as listed in Table S1 of the Supplementary Material [78], indicating that the surface stoichiometry of the films are well maintained for all κ-$Ga_2O_3$ films grown on different substrates.

The bandgap energies of κ-$Ga_2O_3$ films are also estimated by utilizing XPS. By measuring the onset of inelastic loss spectrum on the higher binding energy side of a high intensity peak, the bandgaps of wide bandgap semiconductor materials can be derived [24, 67-70]. The excitation from the valance band to the conduction band is considered to be the lowest energetically inelastic scattering that an electron encounters on its way to the surface. Therefore, the fundamental lower limit of inelastic scattering corresponds to the bandgap of a material [67]. In this study, the strong intensity Ga $2p_{3/2}$ core level spectra is considered for determining the bandgap energies of κ-$Ga_2O_3$ films grown on c-sapphire, GaN- and AlN-on-sapphire and YSZ substrates, as shown in Figures 14(a)-(d). The zoomed image of the background region of the core level is shown in the inset of each figure. The intersection of the linear extrapolation of the loss spectra to the constant background was used to determine the onset of inelastic loss curve. By using the difference between the peak position of Ga $2p_{3/2}$ core level spectra and the onset of inelastic loss spectra, the bandgap energies of κ-$Ga_2O_3$ films grown on c-sapphire, GaN- and AlN-on-sapphire and YSZ substrates are determined to be 5.06 ± 0.25 eV, 5.10 ± 0.20 eV, 5.06 ± 0.20 eV, and 5.08 ± 0.25 eV, respectively, which show a good agreement with the bandgap energies estimated experimentally by optical transmission measurements [40, 48, 52-54]. Further verification of the bandgap values was also carried out using O 1s core-level peak as shown in Figures S2 (a)-(d) of the Supplementary Material [78], which showed consistent $E_g$ values with those obtained from Ga $2p_{3/2}$ core levels.

The band offset at the heterointerfaces between the film and substrate is considered as an important material property for designing and fabricating heterojunction-based devices. In addition to the estimation of bandgap energies of κ-$Ga_2O_3$, we have also determined the band alignments at the interface of κ-$Ga_2O_3$ and various substrates by using XPS. XPS measurements were performed on three types of samples: (a) ~50 nm thick κ-$Ga_2O_3$ films grown on different substrates and (b) bare c-sapphire, GaN- and AlN-on-sapphire templates and YSZ substrates to capture the electronic states from the bulk material and substrates, respectively, and (c) very thin (~2 nm) κ-$Ga_2O_3$ layer grown on top of each substrate to capture the electronic states from the heterointerfaces. The schematic structures of the samples are shown in Figures S3 (a)-(c) of the Supplementary Material [78]. The valence ($\Delta E_v$) band offsets are determined using the Kraut's method as follows [68]:

$$\Delta E_v = (E_{CL_{GaO}}^{GaO} - E_{VBM}^{GaO}) - (E_{CL_{substrate}}^{substrate} - E_{VBM}^{substrate}) - (E_{CL_{Interface}}^{GaO} - E_{CL_{Interface}}^{substrate}) \quad (1)$$

The conduction band offsets ($\Delta E_c$) are calculated using the valance band offsets and bandgap information of κ-$Ga_2O_3$ films and substrates as follows.

$$\Delta E_c = E_g^{substrate} - E_g^{GaO} - \Delta E_v \quad (2)$$

where, $E_{CL_{GaO}}^{GaO}$ and $E_{CL_{substrate}}^{substrate}$ are the binding energies of core levels in bulk materials and substrates, respectively. $E_{CL_{Interface}}^{GaO}$ and $E_{CL_{Interface}}^{substrate}$ represent the binding energies of the core levels for the κ-$Ga_2O_3$ and corresponding substrates determined from the heterostructure between κ-$Ga_2O_3$/substrates. $E_{VBM}^{GaO}$ and $E_{VBM}^{substrate}$ are the positions of valance band maxima (VBM) from κ-$Ga_2O_3$ bulk materials and substrates, respectively. $E_g^{substrate}$ and $E_g^{GaO}$ represent the bandgap energies of the substrates and κ-$Ga_2O_3$ films, respectively. The binding energies of Ga 3d and Ga 3s core levels of κ-$Ga_2O_3$ films grown on all substrates, Al 2p for c-sapphire and AlN-on-sapphire substrates, N 1s for GaN-on-sapphire and Zr 3d for YSZ substrates are considered for valance band offset determination. After applying the Shirley background subtraction, the core level positions are determined by fitting with Voigt (mixed Lorentzian-Gaussian) line shapes. The linear extrapolation of the leading edge of the valance band (VB) spectra to the background is used to determine the VBM position. The fitted Ga 3d and Ga 3s core levels and the VB spectra of κ-$Ga_2O_3$ films grown on GaN-on-sapphire template (Figure 15(a)-(c)), Al 2p core levels for AlN-on-sapphire template (Figure 15(d)) and sapphire substrates (Figure 15(e)), N 1s core level from GaN-on-sapphire template (Figure 15(f)), and Zr 3d core level from YSZ substrates (Figure 15(g)) are shown exemplarily in Figure 15. Ga 3d peak is fitted with three components (Figure 15(a)): O 2s peak at 23.27 eV, Ga-O bond at 20.13 eV and Ga-OH bond at 18.75 eV. Al 2p from AlN/sapphire template are comprised of two sub-peaks (Figure 15(d)): Al-O at 74.54 eV and Al-N bond at lower binding energy of 73.72 eV. The N 1s spectra from GaN/sapphire template can be split into four components (Figure 15(f)): the main peak at 397.5 eV corresponds to Ga-N bond, and the other three peaks at the binding energies of 395.78, 395.23, 392.20 eV are due to the Auger line from Ga [71-74]. The Ga 3s spectra of κ-$Ga_2O_3$ film grown on GaN-on-sapphire template and Al 2p spectra of c-sapphire substrate are fitted with a single peak as shown in Fig. 15 (b) and (e). The Zr 3d spectra from YSZ substrate is decomposed into two sub peaks (Figure 15(g)): Zr $3d_{3/2}$ peak at 184.73 eV and Zr $3d_{5/2}$ at 182.33 eV. Using equation (1), the valance band offsets ($\Delta E_v$) of -0.39, -0.44, -0.28 and -0.32 eV are determined at κ-$Ga_2O_3$/c-sapphire (/GaN, /AlN and /YSZ) heterointerfaces, respectively. The surface roughness can influence the band offset results as XPS is a surface sensitive technique. We observed smooth surface morphologies with low RMS roughness for the κ-$Ga_2O_3$ films grown on c-sapphire, GaN and YSZ substrates as shown in Figs. 4 and 5. However, the rougher surface morphology of κ-$Ga_2O_3$ film grown on AlN-on-sapphire templates may result in uncertainty in the band offset measurement. Note that, the relatively rough interface between κ-$Ga_2O_3$ and AlN, as shown in the high-resolution STEM images in Figure 11, due to the rougher surface morphology of AlN-on-sapphire template (as shown in Fig. 5(c)), strong oxidation of Al adatoms on growth surface and decomposition of AlN under oxygen atmosphere might result in deviation in the valance band offset value extracted from the κ-$Ga_2O_3$/AlN heterostructures. However, the extracted band offset value from the heterointerface does represent the typical material grown by MOCVD. The list of the bandgap energies of κ-$Ga_2O_3$ and native substrates, difference between the positions of different core-level binding energies, and the valance and conduction band offsets are summarized in Table S2 (a)-(d) of Supplementary Material [78] for the heterointerfaces of κ-$Ga_2O_3$/GaN, κ-$Ga_2O_3$/AlN, κ-$Ga_2O_3$/YSZ and κ-$Ga_2O_3$-/c-sapphire. The valance band offsets calculated by using different core levels such as the Ga 3s or Ga 3d (for GaN, AlN and $Al_2O_3$ substrates) and Zr $3d_{3/2}$ or Zr $3d_{5/2}$ (for YSZ substrate) match closely with each other for all κ-$Ga_2O_3$/substrates interfaces.

Finally, the band alignment of κ-$Ga_2O_3$/c-sapphire (/GaN, /AlN and /YSZ) heterointerfaces are shown in Figures 16(a)-(d). The bandgaps of c-sapphire, AlN and YSZ substrates are determined to be 8.59, 6.18 and 5.83 eV by using the onset of inelastic loss spectra of O 1s (c-sapphire and YSZ), and Al 2s (AlN) core levels as shown in Figures S4 (a)-(c) of the Supplementary Material [78]. The bandgap energies estimated for different substrates are found to be well consistent with existing literatures [71-74]. Using equation (2), the conduction band offsets of 3.92, -1.26, 1.40 and 1.07 eV are calculated at κ-$Ga_2O_3$/c-sapphire (/GaN, /AlN and /YSZ) interfaces by using the information of $\Delta E_v$ and bandgap energies of bulk material and substrates. The summary of the valance and conduction band offsets between the interfaces of κ-$Ga_2O_3$ and different native substrates are listed in Table 1. The energy band diagram as shown in Figure 16 shows a type-I (straddling) band discontinuity at the interface of κ-$Ga_2O_3$ and GaN (Fig. 16(b)), whereas type-II (staggered) band alignments are observed at κ-$Ga_2O_3$/c-sapphire (/AlN and /YSZ) heterointerfaces (Figs. 16 (a,c,d)). The type II band alignment in the heterointerfaces is advantageous in unipolar electronic devices as it allows larger band offsets on one side (either conduction or valence band) leading to a strong carrier confinement at the interfaces. Similar to κ-$Ga_2O_3$/AlN (or GaN) interfaces, type-II and type-I band alignments were also reported previously at β-$Ga_2O_3$/AlN [75] and β-$Ga_2O_3$/GaN [76] interfaces, respectively, indicating that both β and κ-polymorphs of $Ga_2O_3$, despite of having different crystal structures, exhibit similar type of band discontinuities with underlying substrates. The conduction band offsets at the interfaces of κ-$Ga_2O_3$ and all investigated substrates are found to be relatively higher than the valance band offsets, indicating great opportunities for excellent electron confinement at the heterointerfaces consisting of κ-$Ga_2O_3$ and substrates. While a sharp interface would enable better carrier confinement with enhanced 2DEG mobility as a result of the abrupt conduction band offset, the scattering due to structural imperfections of the materials including defects, alloy disorder and interface roughness could significantly affect the 2DEG density and mobility. Previously, high 2DEG mobility was achieved in AlGaN/GaN heterostructure with sharp interfaces [77]. In this work, while the sharp interfaces between κ-$Ga_2O_3$ films and c-sapphire, GaN and YSZ substrates show promises for the development of high electron mobility transistors, AlN substrates with smooth surface morphologies are needed to improve the interface quality between κ-$Ga_2O_3$ and AlN to fully utilize its higher conduction band offsets for better carrier confinement.

## IV. Conclusion

In summary, the growth of κ-$Ga_2O_3$ films is successfully demonstrated by MOCVD on different substrates. The physical, structural, and surface morphological properties, chemical homogeneity as well as the interfacial abruptness are investigated as a function of various substrates. Phase pure κ-$Ga_2O_3$ films are grown on YSZ, GaN- and AlN-on-sapphire substrates, although rotational domains are observed in all films owing to the mismatch of rotational symmetry between the substrate and the epilayer. $Ga_2O_3$ films grown on sapphire substrates are found to be composed of both β- and κ- polymorphs. Lower pressure facilitates the growth of phase pure κ-$Ga_2O_3$ films, whereas the optimal growth temperature is found to be 620-650 °C. Relatively rougher surface morphology is observed for κ-$Ga_2O_3$ films grown on AlN-on-sapphire substrates which can be due to the strong oxidation of Al on growth surface. Epitaxial growth of fully relaxed κ-$Ga_2O_3$ layers on sapphire and GaN substrates are observed by RSM mapping. The band gap energy of κ-$Ga_2O_3$ is determined to be ~5.08 eV. A type-I band alignment at κ-$Ga_2O_3$/GaN interface and type-II band alignment at the κ-$Ga_2O_3$/AlN (sapphire and YSZ) interfaces are revealed by the band offset measurements at κ-$Ga_2O_3$/substrates heterointerfaces. While the valance band offsets vary from 0.28 to 0.44 eV, the conduction band offsets for different κ-$Ga_2O_3$/substates are found to be relatively large, ranging from 1.07 to 3.92 eV, indicating a great promise for excellent electron confinement at the interface. The results from this study on the growth of κ-$Ga_2O_3$ thin films and the band offset determined at the heterointerfaces between κ-

$Ga_2O_3$ and substates will provide guidance for future growths, design, and fabrication of κ-$Ga_2O_3$ based power electronic and optoelectronic devices.

## Conflict of Interest Statement

On behalf of all authors, the corresponding author states that there is no conflict of interest.

## Acknowledgements

The authors acknowledge the Air Force Office of Scientific Research FA9550-18-1-0479 (AFOSR, Dr. Ali Sayir) for financial support. The authors also acknowledge the National Science Foundation (Grant No. 1810041, No. 2019753) and Semiconductor Research Corporation (SRC) under the Task ID GRC 3007.001 for partial support. Electron microscopy was performed at the Center for Electron Microscopy and Analysis (CEMAS) at The Ohio State University.

## Data Availability

The data that support the findings of this study are available from the corresponding author upon reasonable request.

## Table Caption

**Table 1.** Summary of the valance and conduction band offsets at κ-$Ga_2O_3$/GaN, κ-$Ga_2O_3$/AlN, κ-$Ga_2O_3$/YSZ, and κ-$Ga_2O_3$/sapphire heterointerfaces, determined by using different core level peaks from XPS spectra.

## Figure Captions

**Figure 1.** XRD ω-2θ scan profiles for the (004) reflections of κ-$Ga_2O_3$ films grown on (a) YSZ, (b) AlN- and (c) GaN-on-sapphire templates and (d) c-sapphire substrates.

**Figure 2.** X-ray asymmetrical reciprocal space maps around (139) κ-$Ga_2O_3$ grown on (a) c-sapphire and (b) GaN-on-sapphire substrates. $(11\bar{2}.12)$ and $(10\bar{1}5)$ reflections are originated from sapphire and GaN substrates, respectively.

**Figure 3.** XRD Φ-scan profiles of {139} κ-$Ga_2O_3$ with (a) $\{10\bar{1}2\}$ $Al_2O_3$, $\{10\bar{1}5\}$ reflections of (b) GaN and (c) AlN as well as (d) {311} reflection from YSZ substates.

**Figure 4.** Surface view FESEM images of κ-$Ga_2O_3$ films grown on (a) c-sapphire, (b) GaN- and (c) AlN-on-sapphire templates and (d) YSZ substrates.

**Figure 5.** Surface AFM images of (a) c-sapphire, (b) GaN-on-sapphire, (c) AlN-on-sapphire and (d) YSZ bare substrates. AFM images of κ-$Ga_2O_3$ films grown on (e, i, m) c-sapphire, (f, j, n) GaN- and (g, k, o) AlN-on-sapphire templates and (h. l, p) YSZ substrates with TEGa flow of (e-h) 80 sccm, (i-l) 130 sccm and (m-p) 220 sccm.

**Figure 6.** Surface AFM images of κ-$Ga_2O_3$ films grown with (a-d) 20 nm, (e-h) 50 nm and (i - l) 194 nm thickness on (a, e, i) c-sapphire, (b, f, j) GaN- and (c, g, k) AlN-on-sapphire templates and (d. h, l) YSZ substrates with same TEGa flow of 220 sccm.

**Figure 7.** XRD ω-2θ scan spectra of $Ga_2O_3$ films grown on c-sapphire substrate at a chamber pressure of (a) 20 torr and (b) 7 torr with the same growth temperature of 650 °C.

**Figure 8.** XRD ω-2θ scan spectra of κ-$Ga_2O_3$ films grown on (a) c-sapphire, (b) GaN- and (c) AlN-on-sapphire templates and (d) YSZ substrates at growth temperature of 510, 550 and 620 °C with the same chamber pressure of 5 torr.

**Figure 9.** (a) HAADF STEM image of κ-$Ga_2O_3$ thin films grown on c-sapphire at growth temperature of 650 °C, chamber pressure of 7 torr and TEGa flow rate of 80 sccm. Two diffraction patterns shown (top and bottom right) were acquired from the area (i) and (ii), respectively. The inset in each diffraction pattern shows the matching simulated diffraction pattern. (b) LAADF STEM image of the same area. Diffraction pattern shows the experimental and simulated (inset) diffraction patterns from the c-sapphire substrate. High magnification images of the (c) interface, and (d) the selected regions near surface the thin films. The (c) inset shows the schematic of the orthorhombic structure with three different rotation domains e.g. (100), $(\bar{1}10)$ and $(1\bar{1}0)$.

**Figure 10.** STEM data showing (a) HAADF and (b) LAADF images along with experimental and simulated (inset) diffraction patterns. (c) High magnification image of the interface of κ-$Ga_2O_3$ thin films grown on (0001) GaN grown at growth temperature of 620 °C, chamber pressure of 5 torr and TEGa flow rate of 220 sccm.

**Figure 11.** STEM data showing (a) HAADF and (b) LAADF images along with experimental and simulated (inset) diffraction patterns. (c) High magnification image of the interface of κ-$Ga_2O_3$ thin films grown on (0001) AlN at growth temperature of 620 °C, chamber pressure of 5 torr and TEGa flow rate of 220 sccm.

**Figure 12.** STEM data showing (a) HAADF and (b) LAADF images along with experimental and simulated (inset) diffraction patterns. (c) High magnification image of the interface of κ-$Ga_2O_3$ thin films grown on (100) YSZ at growth temperature of 620 °C, chamber pressure of 5 torr and TEGa flow rate of 220 sccm, and (d) selected regions with atomic arrangement resolved along [010] crystal orientation.

**Figure 13.** (a) XPS survey spectra of κ-$Ga_2O_3$ films grown on c-sapphire, GaN- and AlN-on-sapphire templates and YSZ substrates. (b) Ga 3s and (c) O 1s core level spectra for κ-$Ga_2O_3$ film grown on YSZ substrate. Experimental data points (black open circles) are fitted using mixed Lorentzian-Gaussian line shapes (black solid lines) after applying the Shirley background (gray solid lines).

**Figure 14.** The bandgap energies of κ-$Ga_2O_3$ films grown on (a) c-sapphire, (b) GaN- and (c) AlN-on-sapphire templates and (d) YSZ substrates determined by analyzing the energy difference between Ga $2p_{3/2}$ peak position and the onset of energy loss spectrum. The insets show the zoomed view of the background region of the Ga $2p_{3/2}$ core levels.

**Figure 15.** Fitted (a) Ga 3d, (b) Ga 3s and (c) valance band (VB) spectra of κ-$Ga_2O_3$ film grown on GaN-on-sapphire template. The Al 2p core levels from (d) AlN-on-sapphire and (e) c-sapphire substrates, (f) N 1s from GaN-on-sapphire template, and (g) Zr 3d from YSZ substrates are also included. Experimental data are shown as black open circles and the fitted curves (envelope) are represented as black sloid lines.

**Figure 16.** Summary of the valance and conduction band offsets at the interface of κ-$Ga_2O_3$ and (a) c-sapphire, (b) GaN, (c) AlN, and (d) YSZ substrates.

**Table 1.**

| Core levels | Valance band offsets, $\Delta E_v$ (eV) | | | | Conduction band offsets, $\Delta E_c$ (eV) | | | |
|---|---|---|---|---|---|---|---|---|
| | $\kappa$-$Ga_2O_3$ /GaN | $\kappa$-$Ga_2O_3$ /AlN | $\kappa$-$Ga_2O_3$ /YSZ | $\kappa$-$Ga_2O_3$/ sapphire | $\kappa$-$Ga_2O_3$ /GaN | $\kappa$-$Ga_2O_3$ /AlN | $\kappa$-$Ga_2O_3$ /YSZ | $\kappa$-$Ga_2O_3$/ sapphire |
| Ga 3s, N 1s | -0.49 | -- | -- | -- | -1.21 | -- | -- | -- |
| Ga 3d, N 1s | -0.44 | -- | -- | -- | -1.26 | -- | -- | -- |
| Ga 3s, Al 2p | -- | -0.28 | -- | -0.49 | -- | 1.40 | -- | 4.02 |
| Ga 3d, Al 2p | -- | -0.28 | -- | -0.39 | -- | 1.40 | -- | 3.92 |
| Ga 3d, Zr $3d_{3/2}$ | -- | -- | -0.32 | -- | -- | -- | 1.07 | -- |
| Ga 3d, Zr $3d_{5/2}$ | -- | -- | -0.29 | -- | -- | -- | 1.04 | -- |

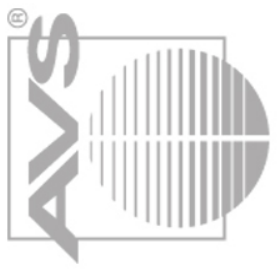

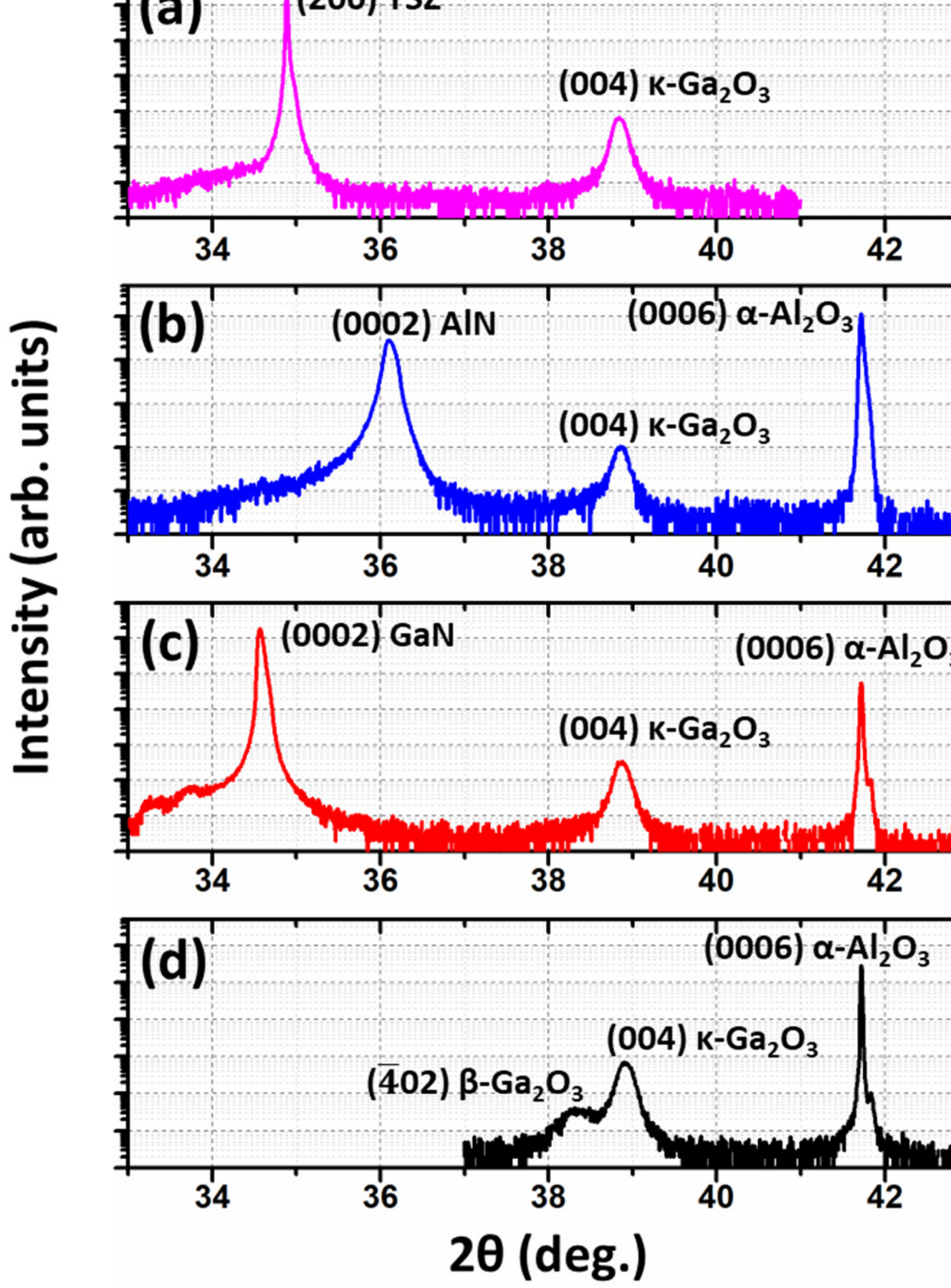

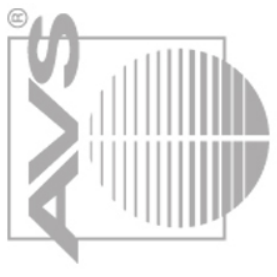

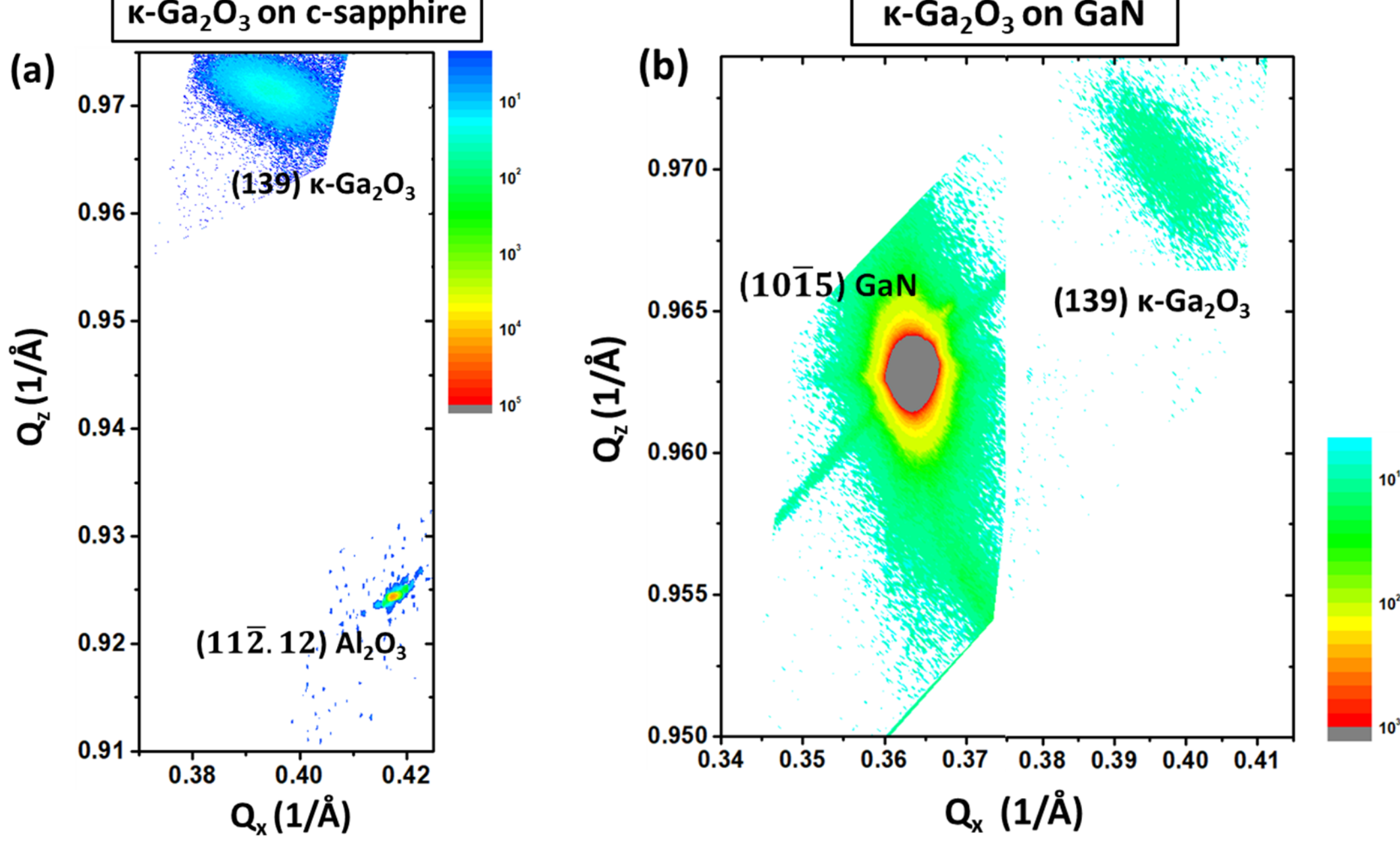

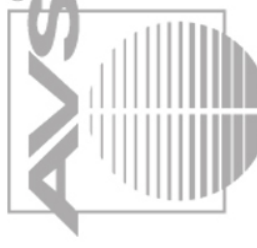

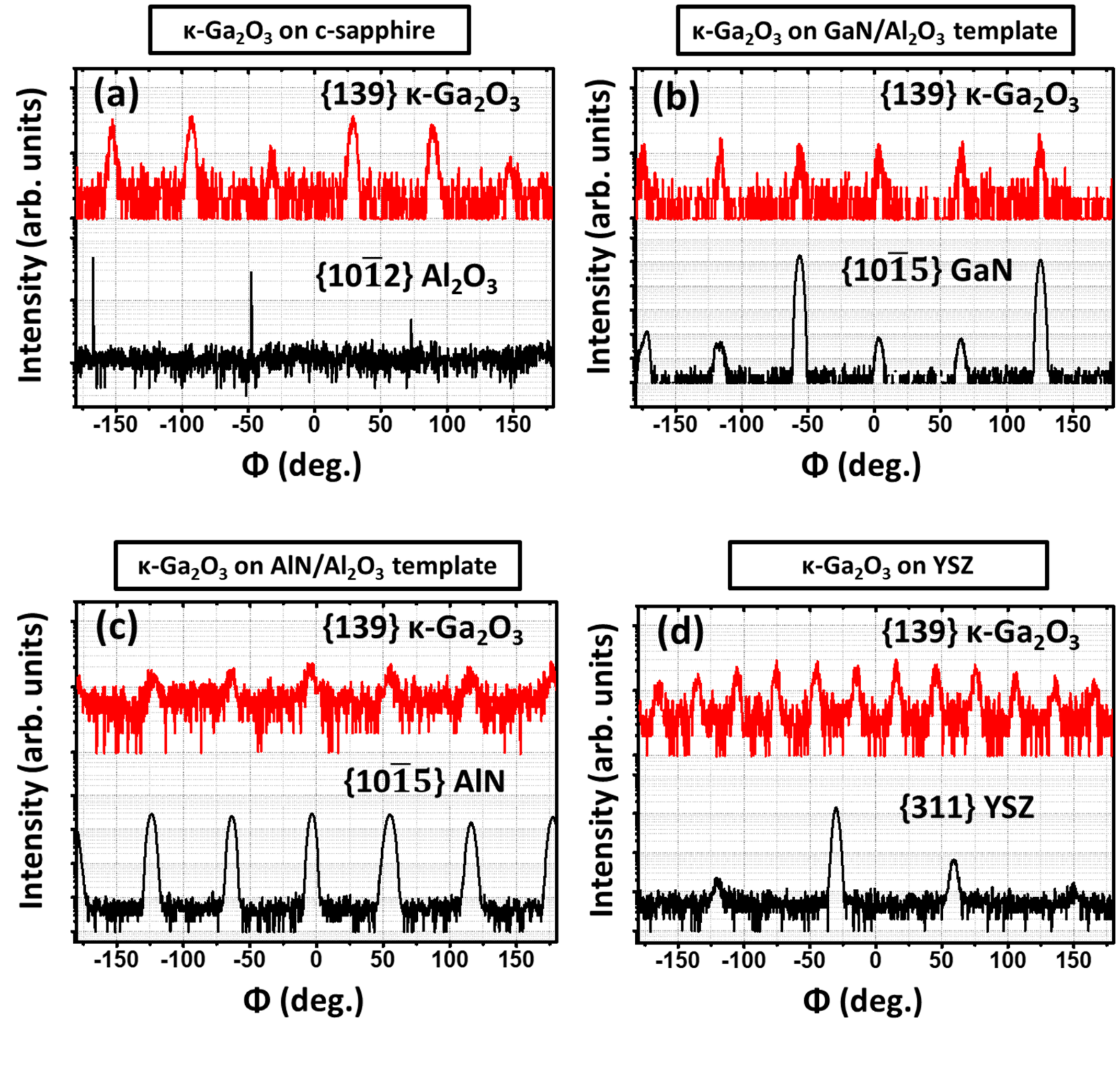

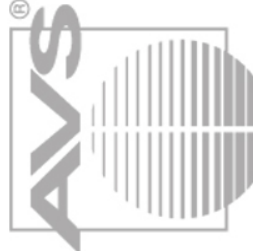

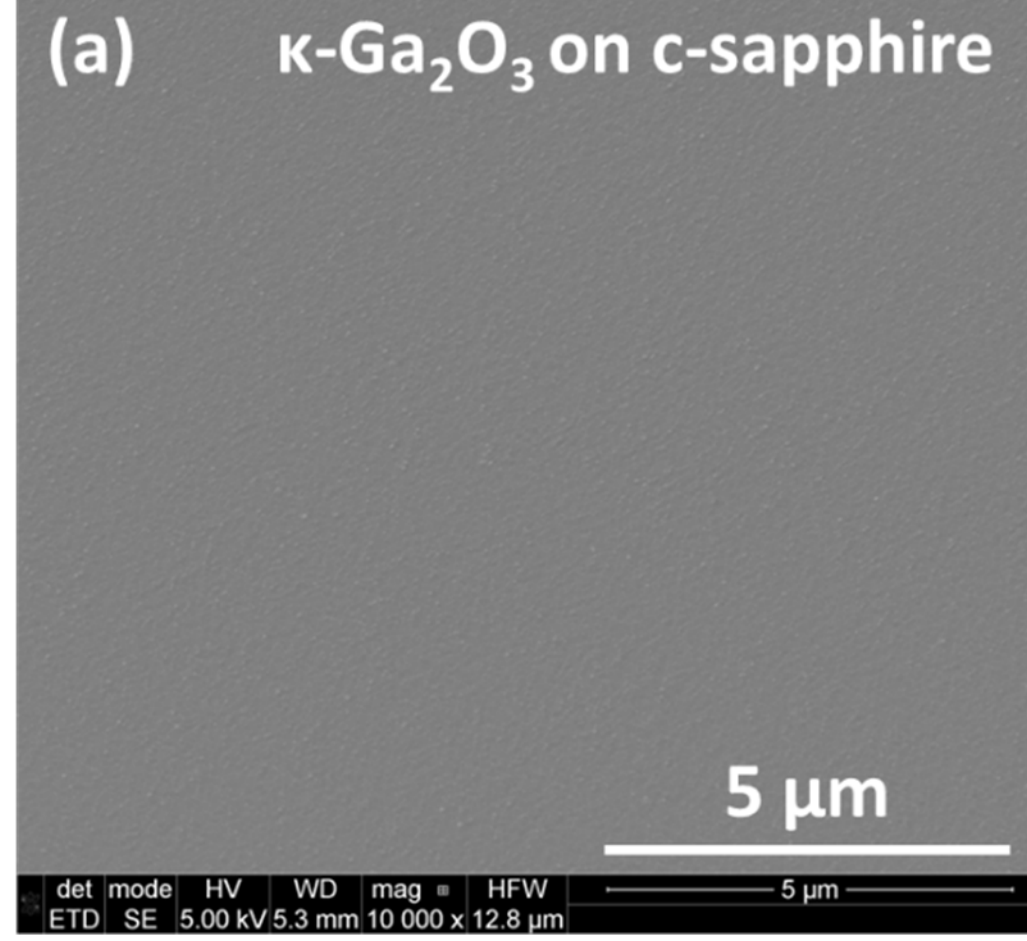

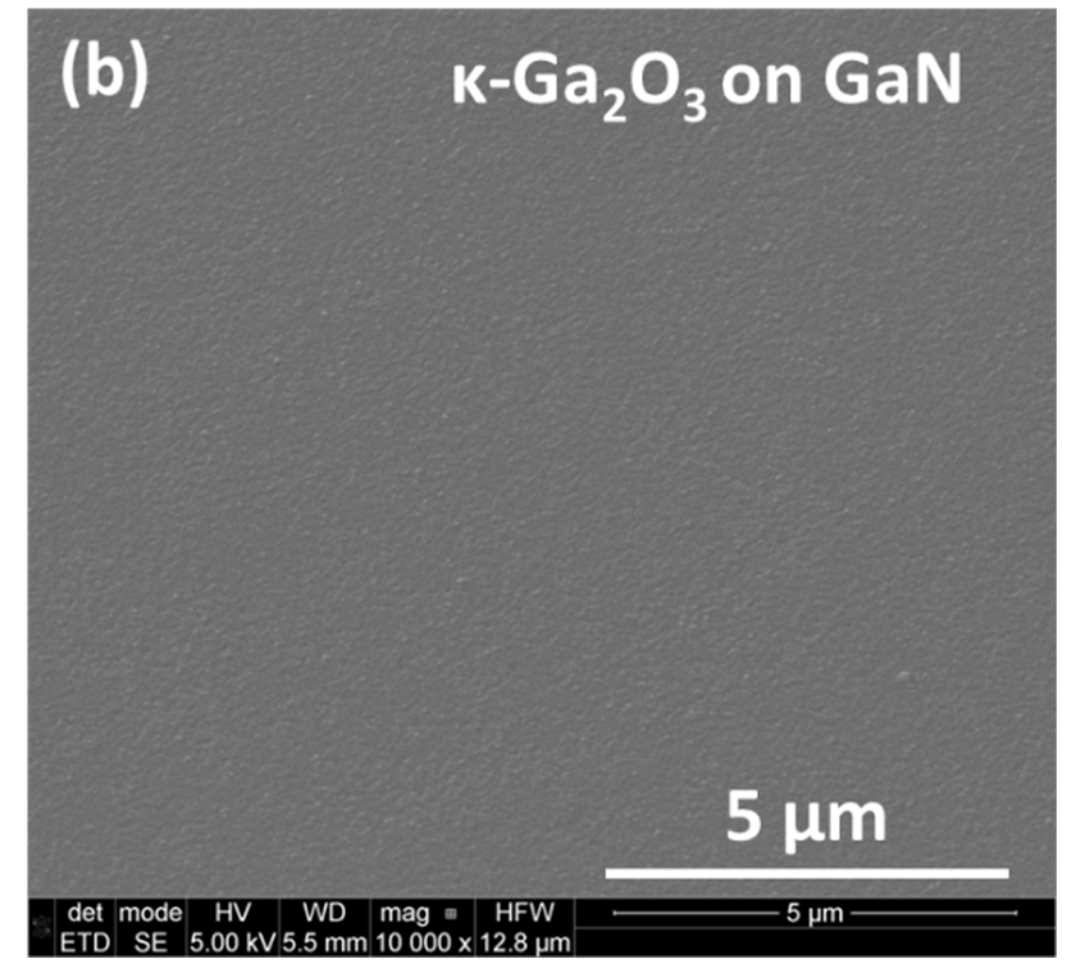

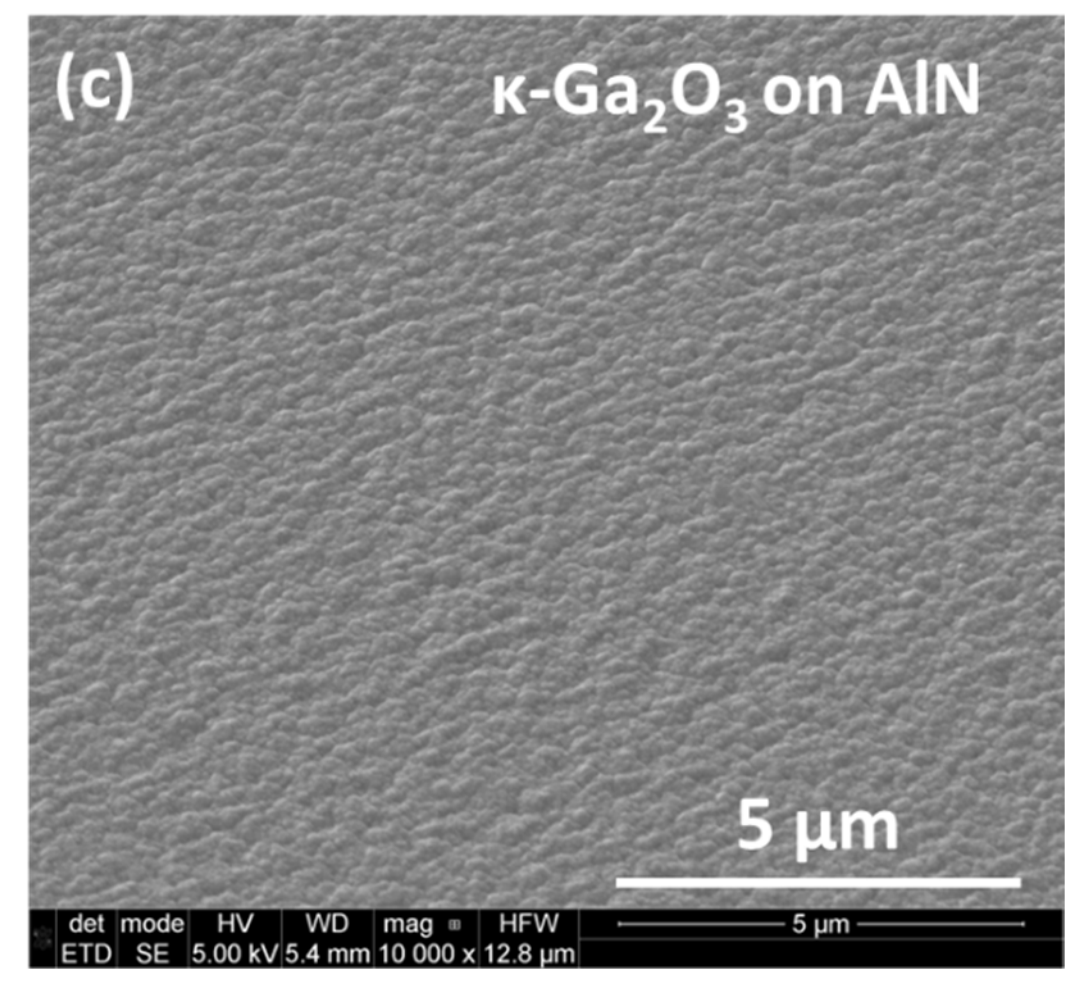

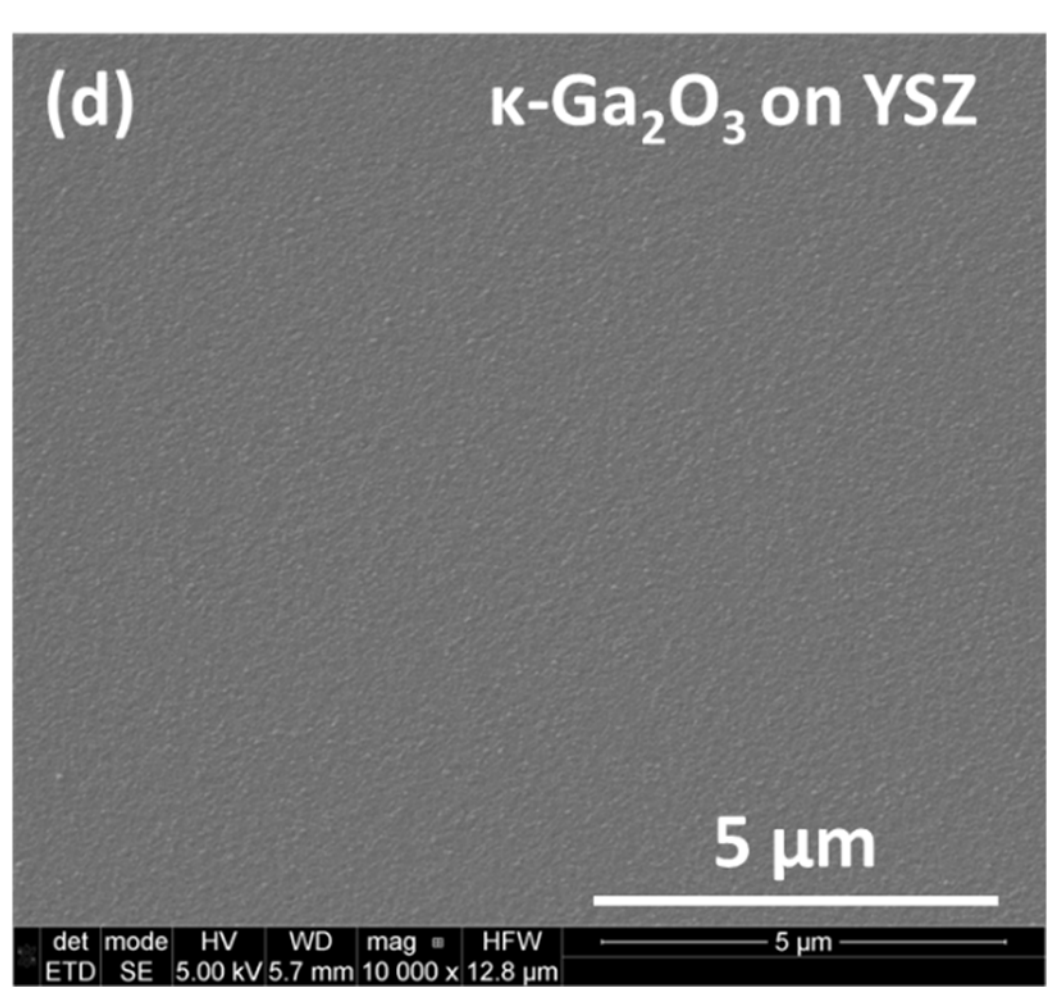

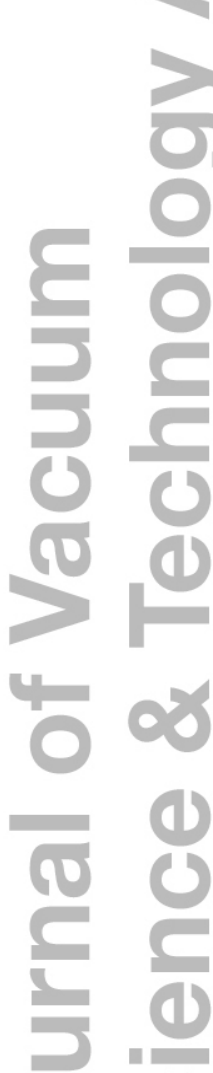

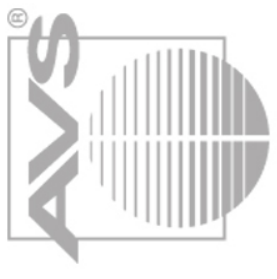

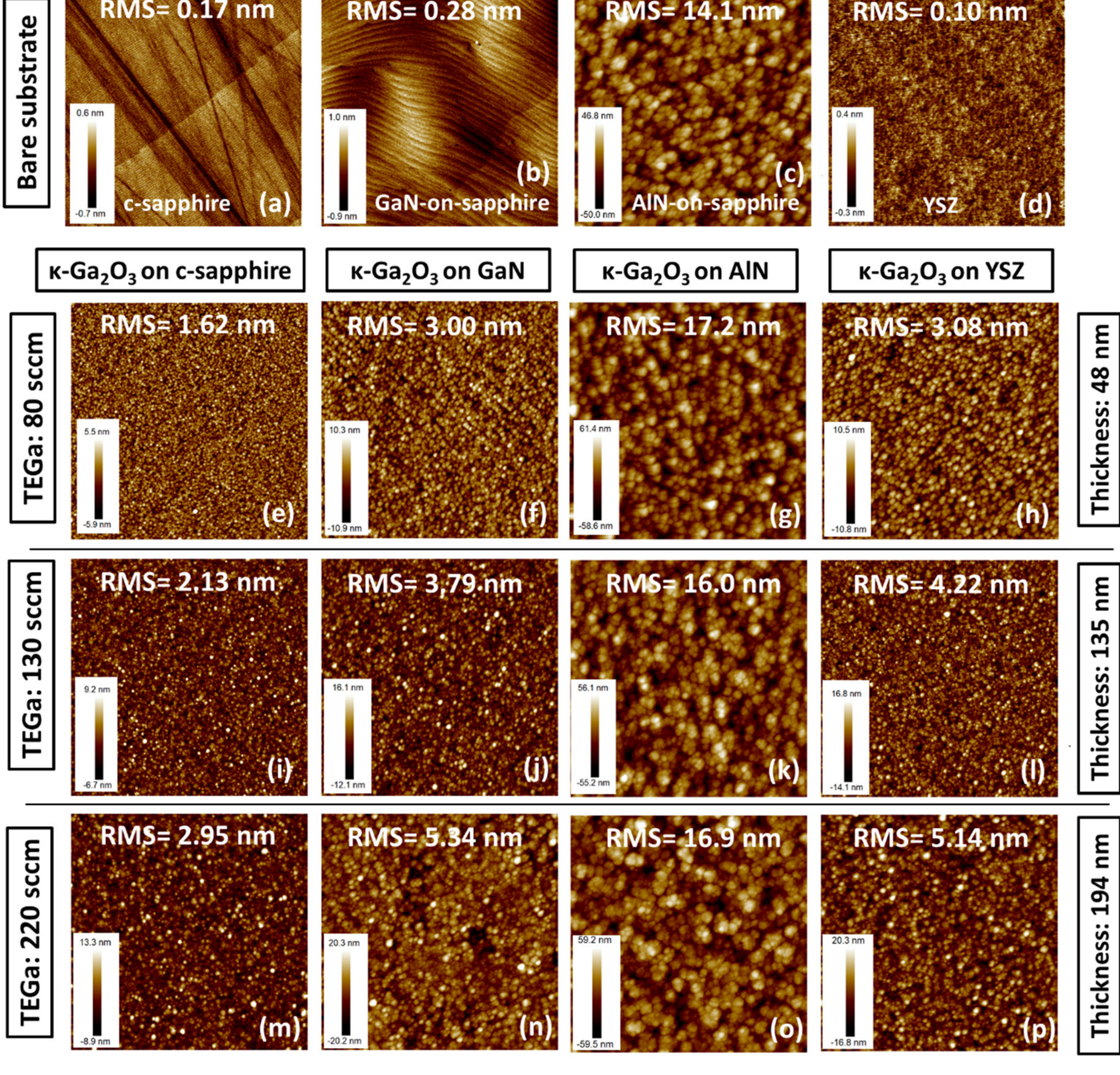

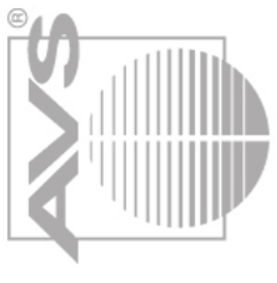

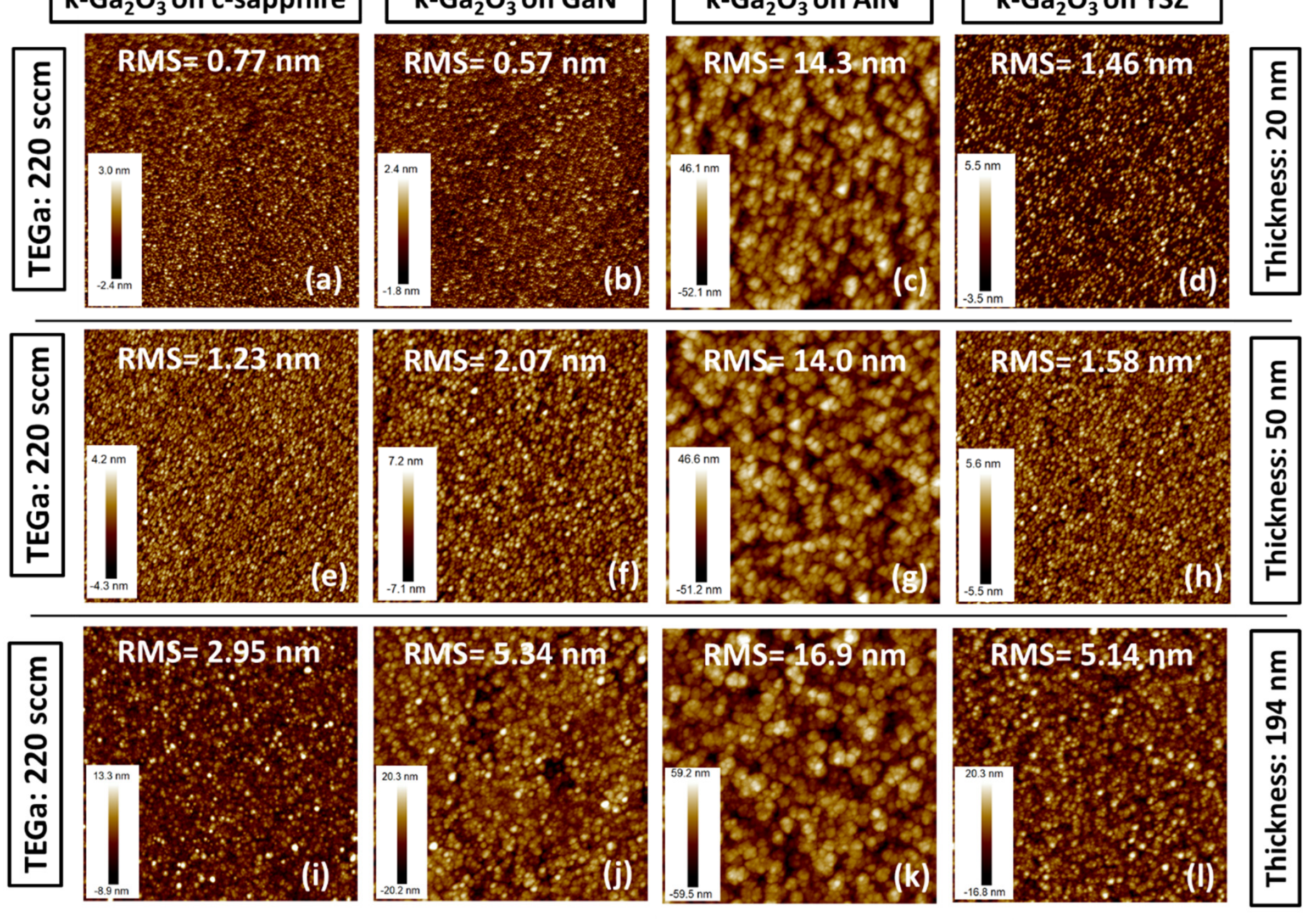

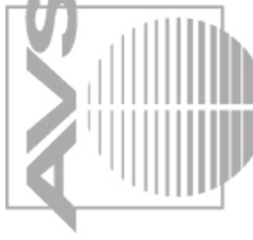

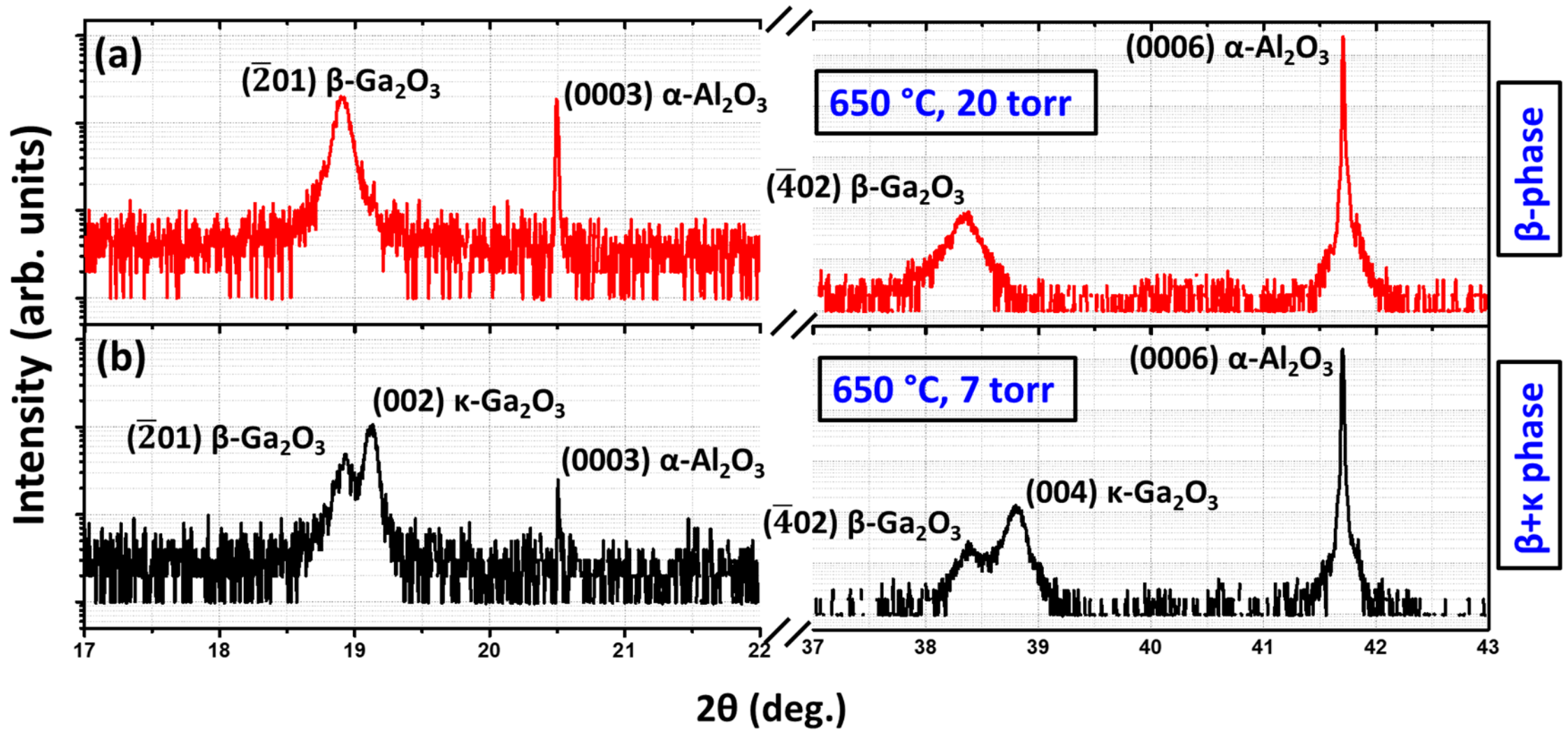

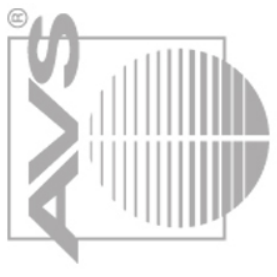

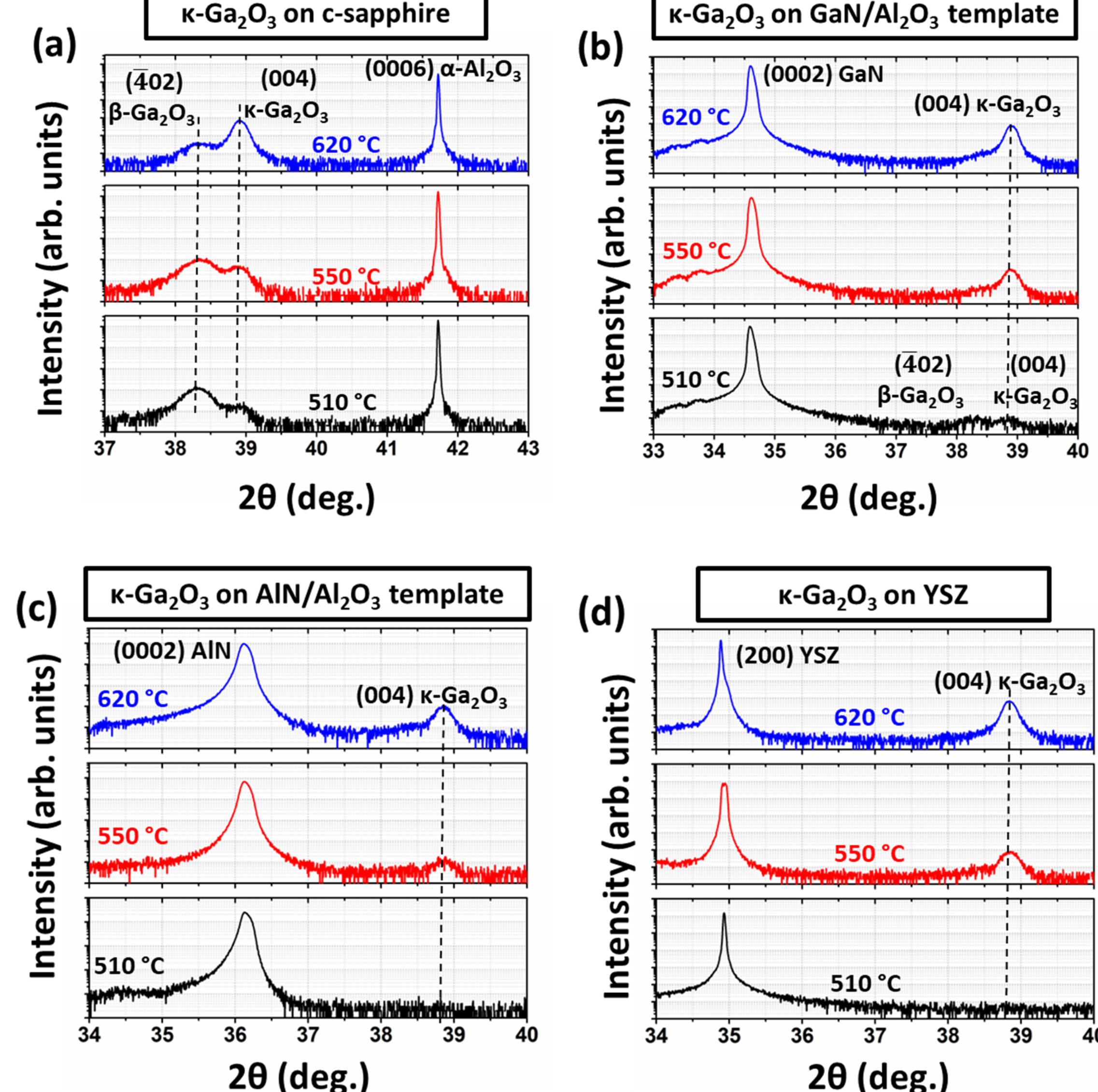

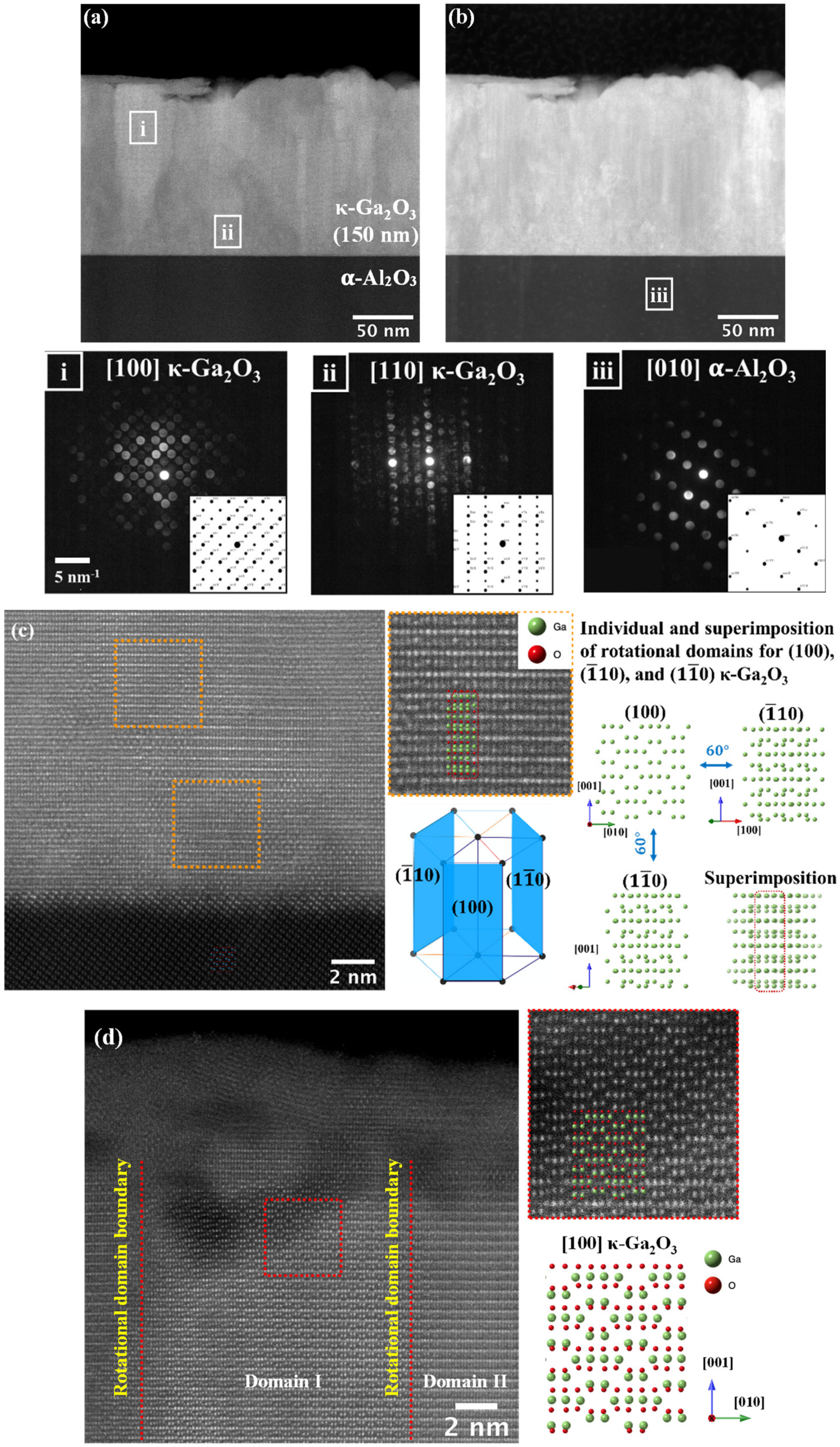

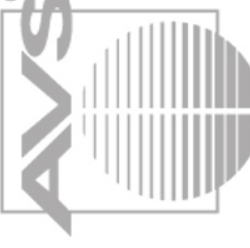

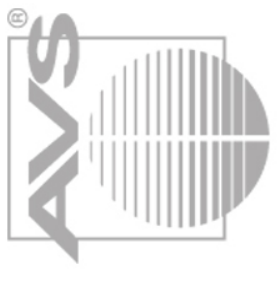

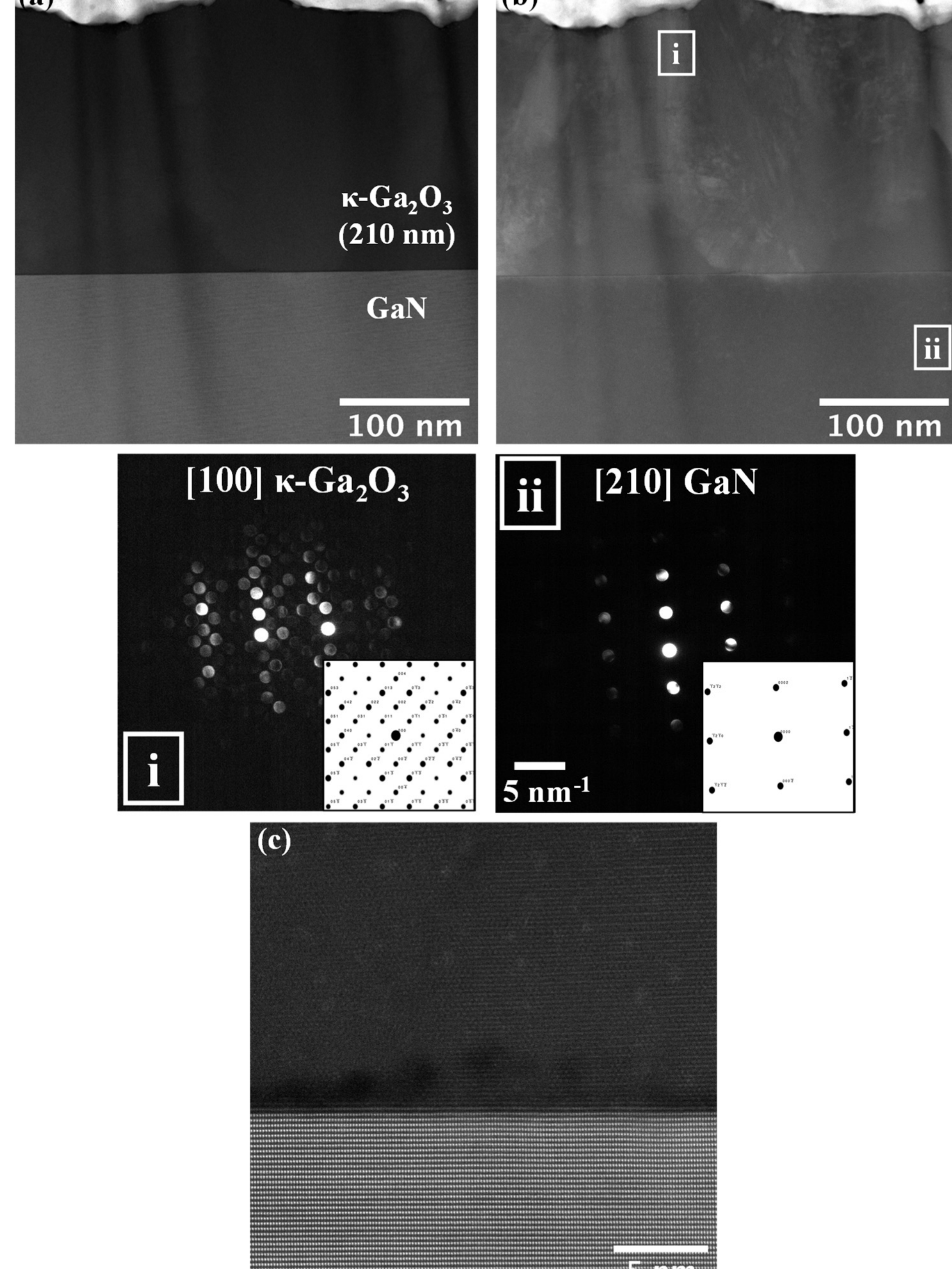

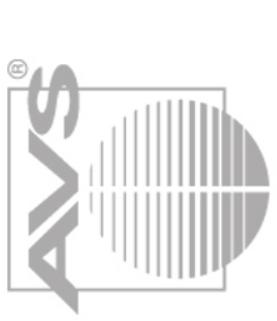

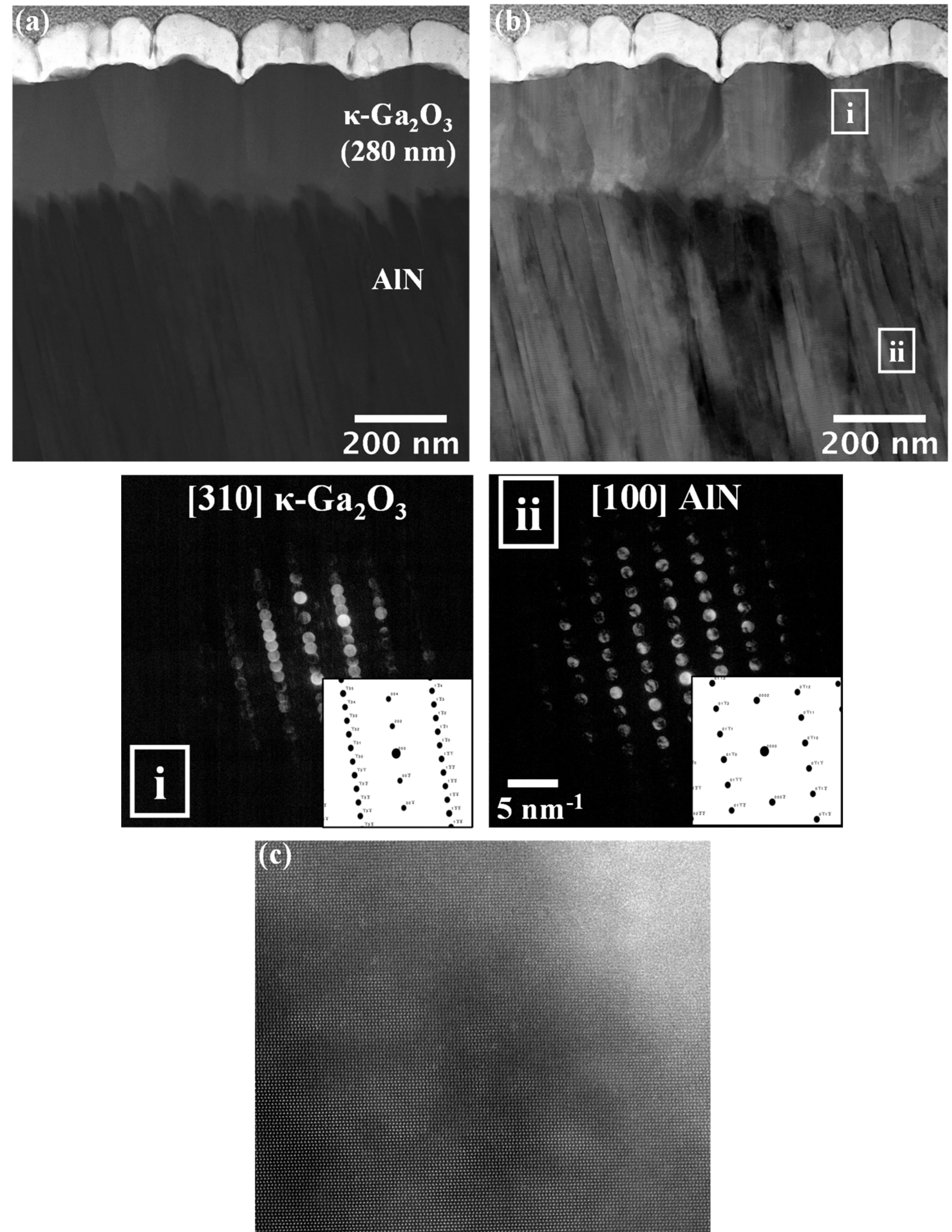

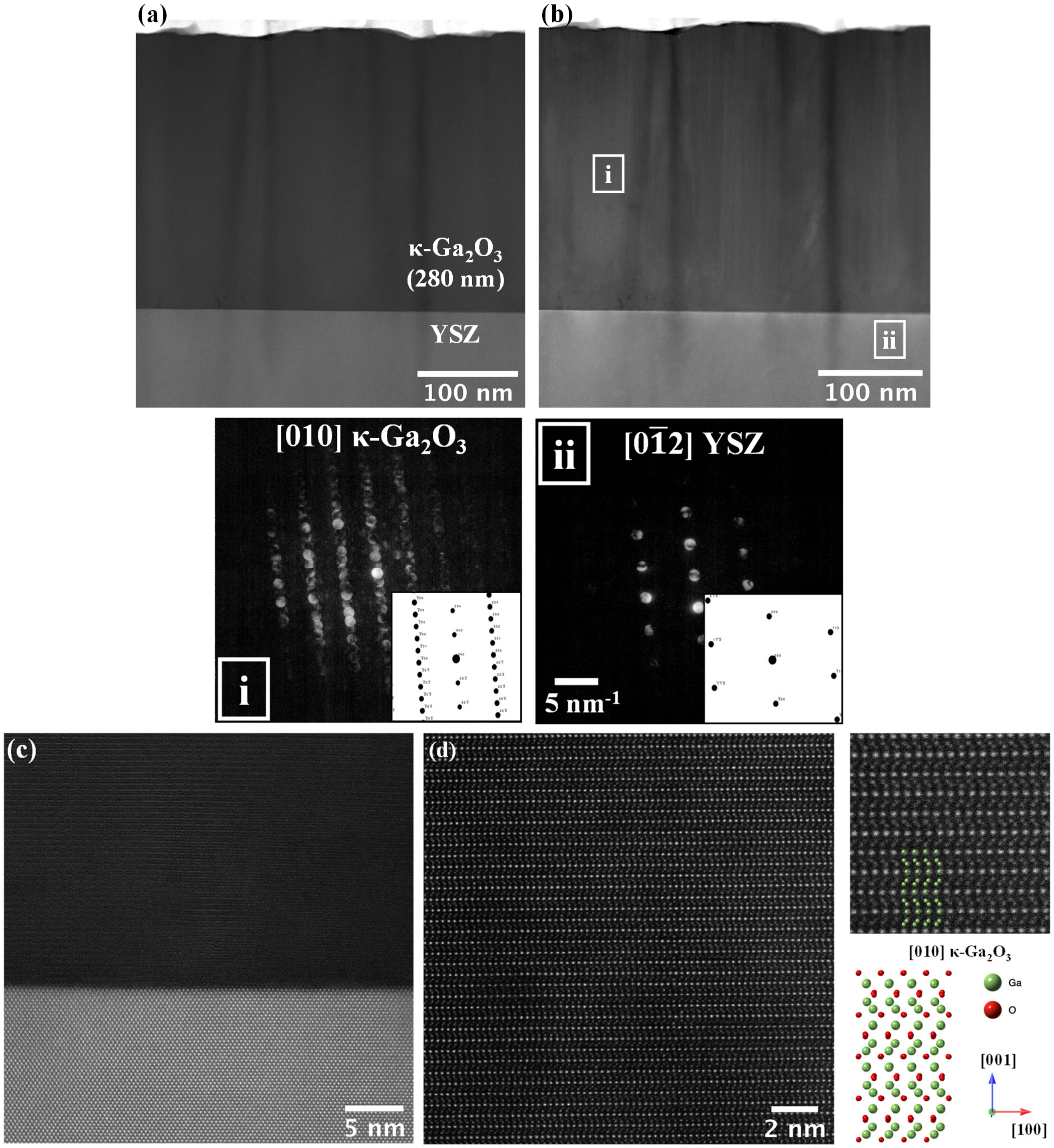

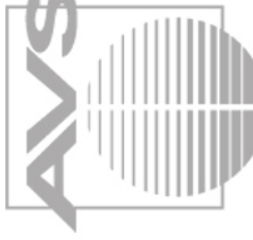

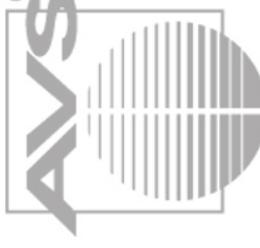

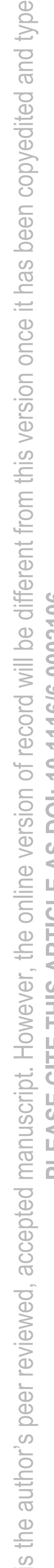

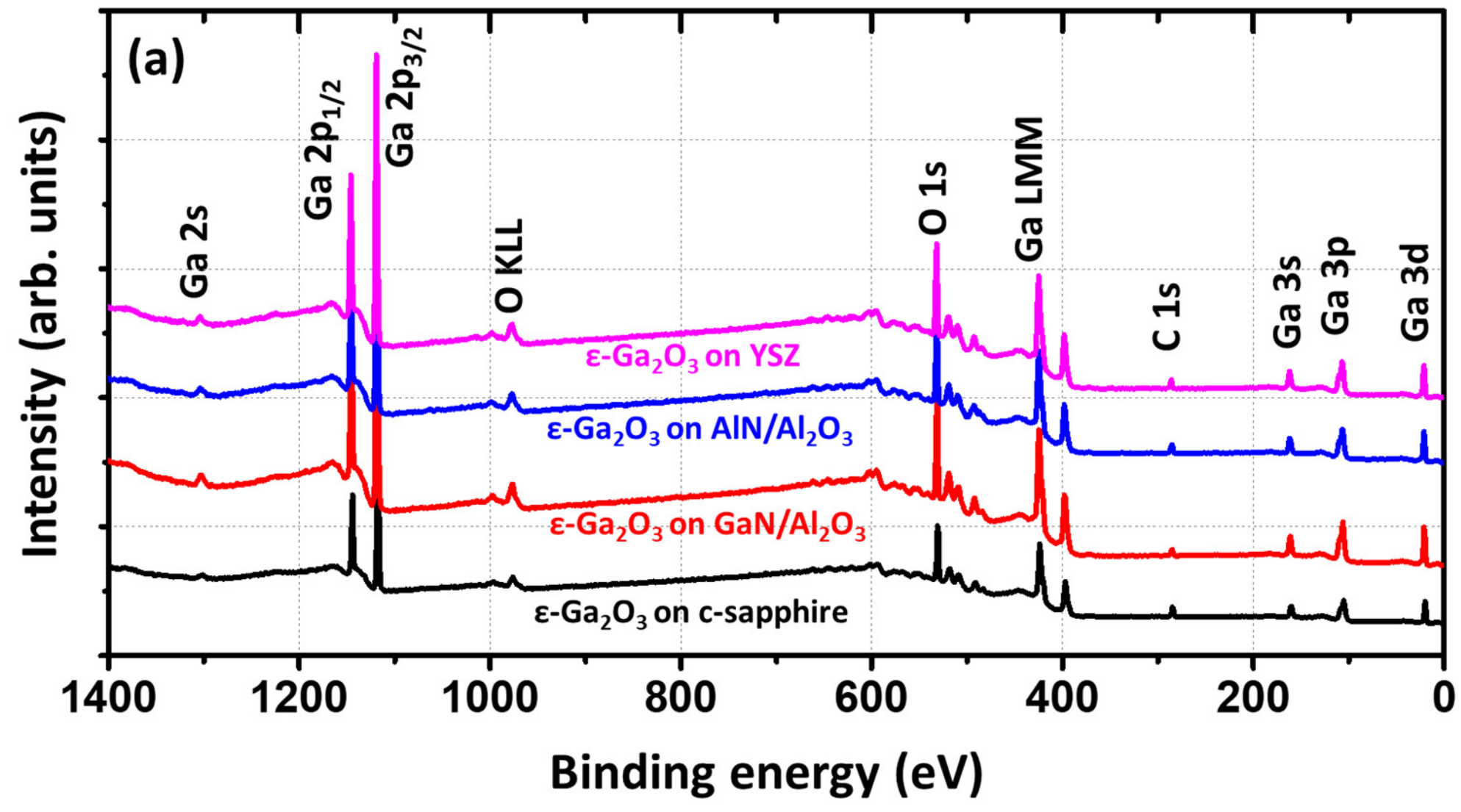

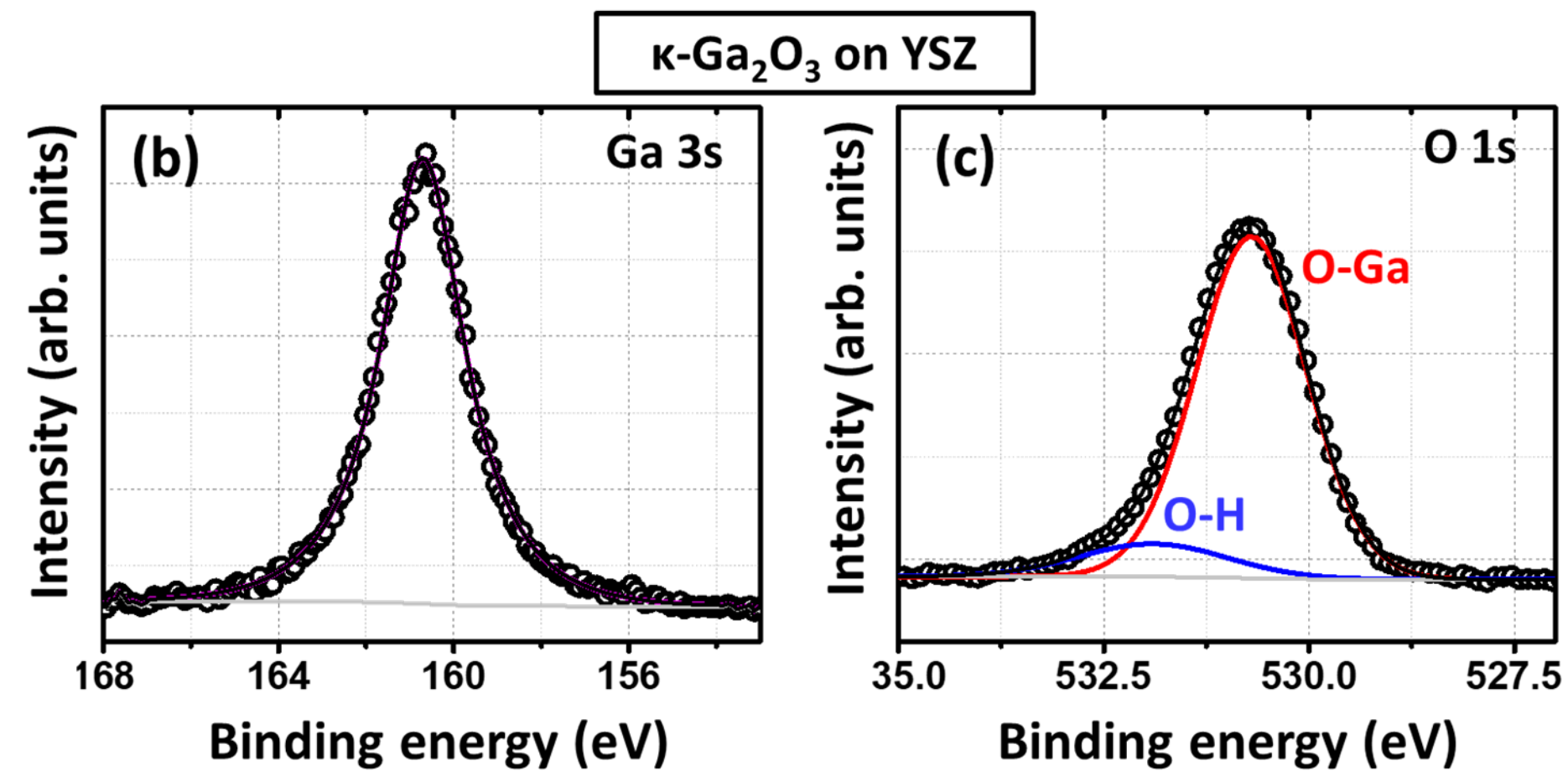

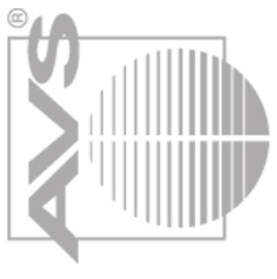

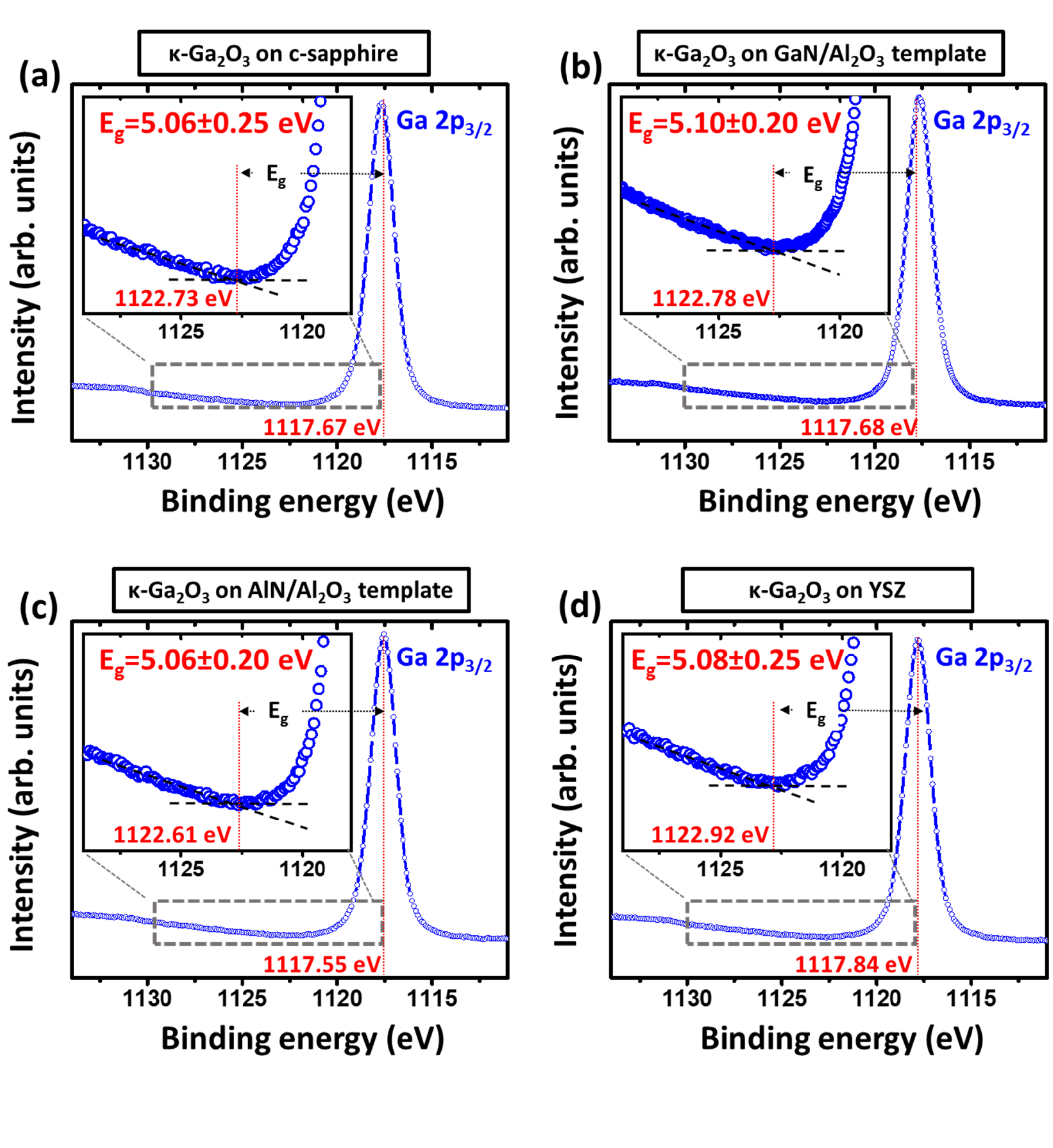

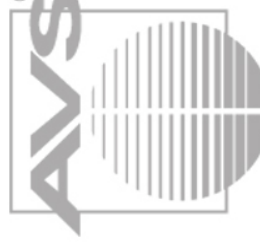

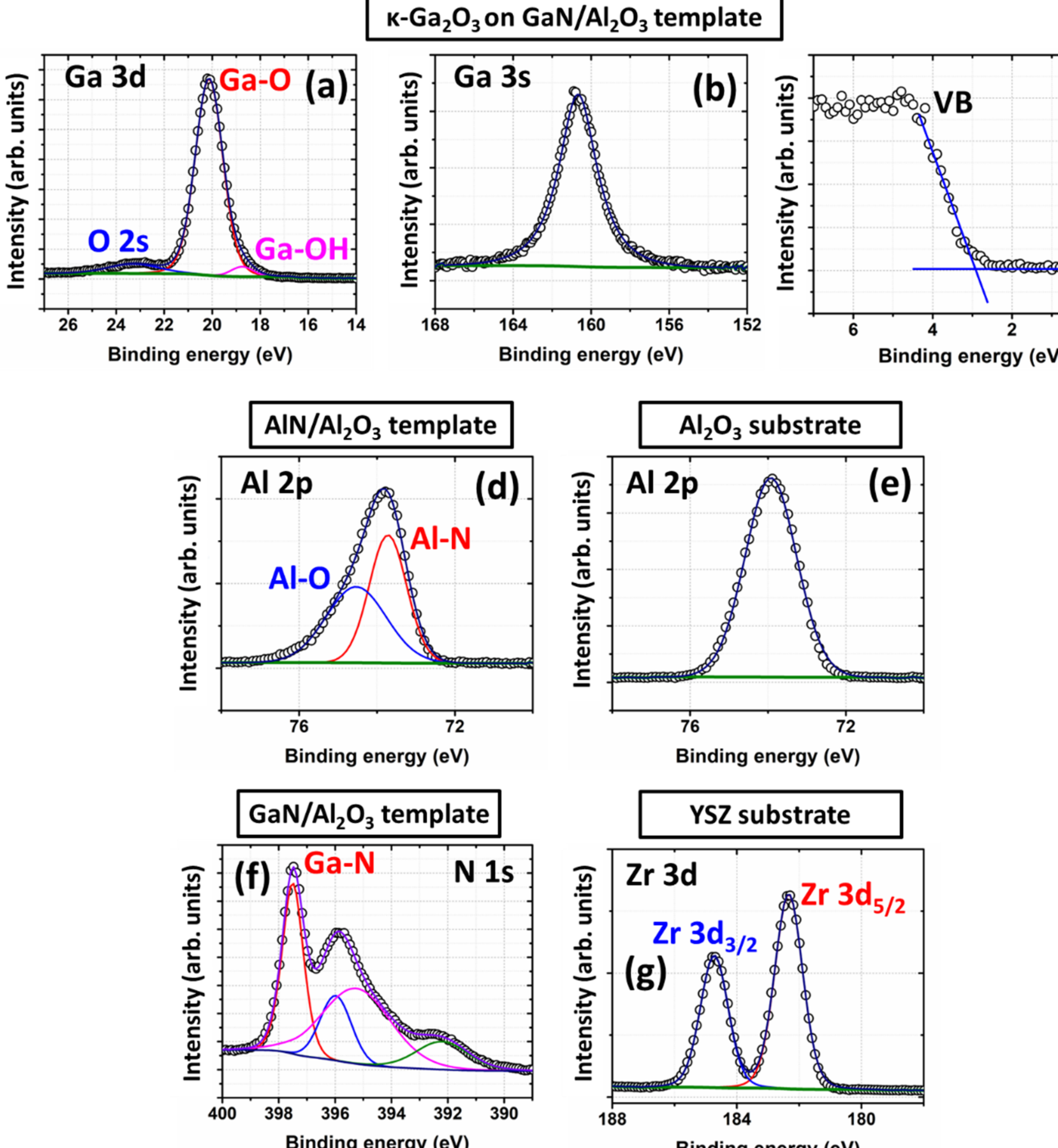

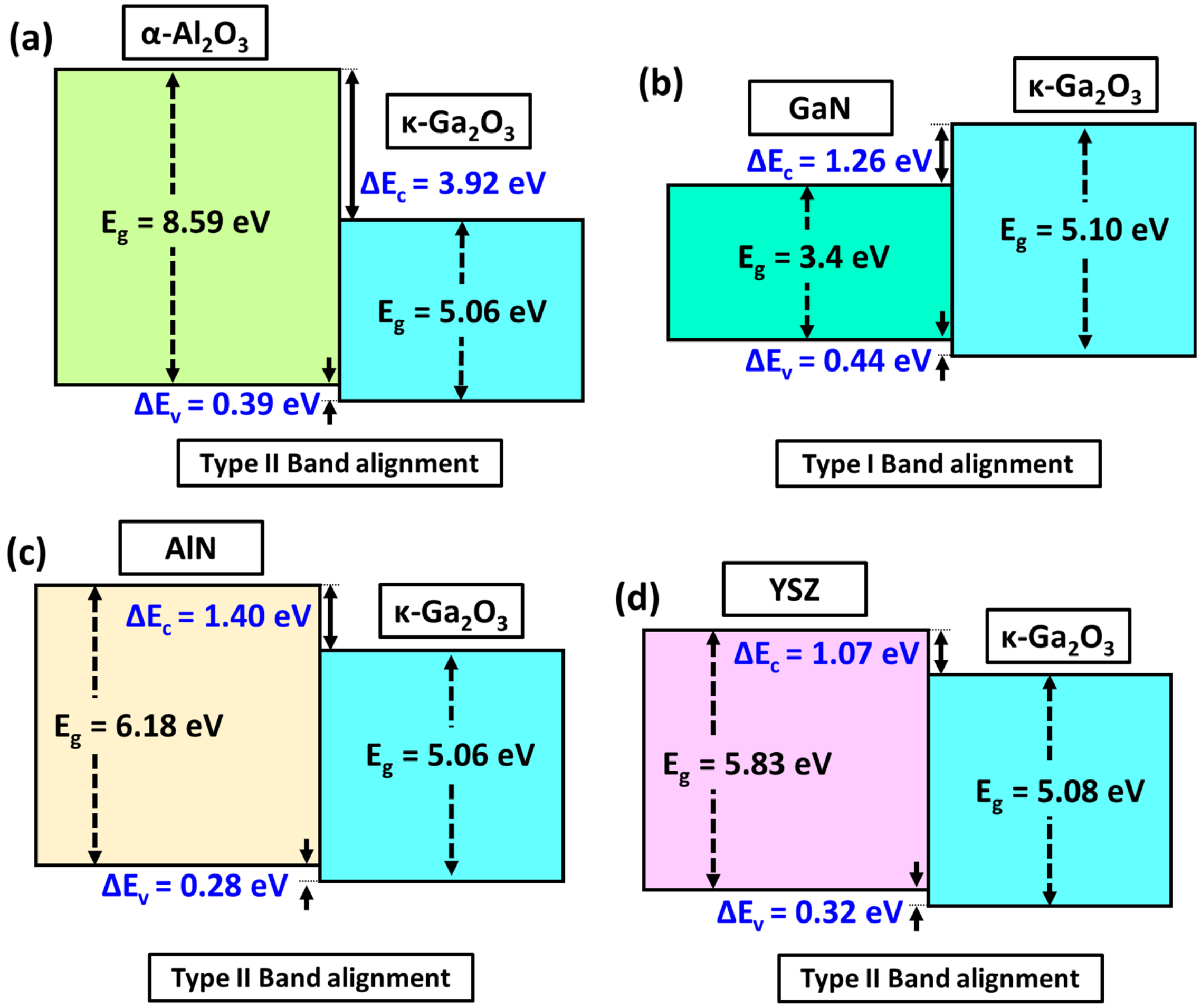

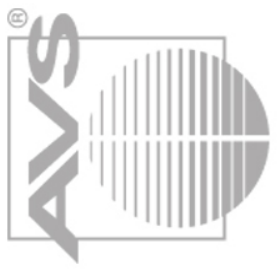

(a)

α-$Al_2O_3$

κ-$Ga_2O_3$

$E_g$ = 8.59 eV

$\Delta E_c$ = 3.92 eV

$E_g$ = 5.06 eV

$\Delta E_v$ = 0.39 eV

Type II Band alignment

(b)

GaN

κ-$Ga_2O_3$

$\Delta E_c$ = 1.26 eV

$E_g$ = 3.4 eV

$E_g$ = 5.10 eV

$\Delta E_v$ = 0.44 eV

Type I Band alignment

(c)

AlN

κ-$Ga_2O_3$

$\Delta E_c$ = 1.40 eV

$E_g$ = 6.18 eV

$E_g$ = 5.06 eV

$\Delta E_v$ = 0.28 eV

Type II Band alignment

(d)

YSZ

κ-$Ga_2O_3$

$\Delta E_c$ = 1.07 eV

$E_g$ = 5.83 eV

$E_g$ = 5.08 eV

$\Delta E_v$ = 0.32 eV

Type II Band alignment